\documentclass[
 reprint,
 amsmath,amssymb,
 nofootinbib,
 twocolumn,
 apj,
 superscriptaddress,
 floatfix
]{aastex701}

\usepackage{booktabs}
\begin{document}

\title{Gravitational-Wave Microlensing as a Probe of Compact Object Population}

\author[orcid=0000-0002-8588-4794]{Eungwang Seo}
\thanks{e.seo.1@research.gla.ac.uk}
\affiliation{SUPA, School of Physics and Astronomy, University of Glasgow, Glasgow G12 8QQ, United Kingdom}
\email{e.seo.1@research.gla.ac.uk}

\author[orcid=0000-0002-7213-3211]{Jose Mar\'ia Ezquiaga} 
\affiliation{Center of Gravity, Niels Bohr Institute, Blegdamsvej 17, 2100 Copenhagen, Denmark}
\email{}

\author[0000-0001-8322-5405]{Martin Hendry}
\affiliation{SUPA, School of Physics and Astronomy, University of Glasgow, Glasgow G12 8QQ, United Kingdom}
\email{}

\begin{abstract}
The abundance of compact objects ranging from stellar to intermediate masses ($\sim 1 - 10^{5}M_{\odot}$) remains poorly constrained, particularly for populations such as primordial black holes and dark matter subhalos that emit little or no electromagnetic radiation.
Gravitational-wave (GW) microlensing provides a complementary way to probe these populations through wave-optics effects imprinted on the GW signals, with implications for astrophysical and cosmological inference.
We develop a hierarchical Bayesian framework to infer the population properties of microlenses from an ensemble of microlensed GW events, constraining the slope of the power-law lens mass function ($\alpha_{\rm ML}$), the lens number density ($n_0$), and the fraction of dark matter in compact lenses ($f_{\rm DM}$).
Using simulated microlensed GW signals drawn from a realistic binary black hole population, we find that $\sim 100$ detected microlensed events are able to constrain both $\alpha_{\rm ML}$ and $n_0$, although a degeneracy between the two remains.
For a low-mass-dominated lens population ($\alpha_{\rm ML}=2$) with $n_0=10^6~\mathrm{Mpc}^{-3}$, a three-year observation at the O4 sensitivity yields $\sim 130$ detectable microlensed events and recovers the injected dark-matter fraction, $f_{\rm DM}\simeq 1.2 \times 10^{-4}$, with a 90\% credible interval of $7.3 \times 10^{-5} \lesssim f_{\rm DM} \lesssim 1.8 \times 10^{-4}$.
\end{abstract}

\section{Introduction}\label{sec:intro}

The abundance and mass distribution of dark compact objects are difficult to determine as they may produce little or no electromagnetic (EM) emission~\citep{cardoso2019testing}.
Such objects may include astrophysical black holes~\citep{belczynski2004comprehensive,coleman2004intermediate,narayan2005black,2026arXiv260527226T}, primordial black holes (PBHs)~\citep{carr2016primordial,sasaki2018primordial,green2021primordial,carr2025history}, dark-matter (DM) clumps~\citep{silk1993clumpy,diemand2008clumps,buckley2018collapsed}, and more exotic compact objects~\citep{d2018massive,chang2019structure,ryan2022exotic,zhang2023dark,bramante2024effect}.

Gravitational microlensing provides a powerful way to probe such otherwise elusive populations that could constitute a fraction of DM~\citep{alcock2001macho,fedorova2016gravitational,zumalacarregui2018limits,oguri2018understanding,niikura2019microlensing,mroz2019microlensing,abrams2020hunting,winch2022using,blaineau2022new,fardeen2024astrometric,perkins2024disentangling,romao2025dark,green2026stellar}.
Rather than requiring the lens to be directly observable, microlensing relies on the gravitational imprint of the lens on the source signal.
The resulting lensing effect depends on the mass distribution of the lens, the relative geometry between the lens and source, and the evolution of this geometry due to their relative motion~\citep{schneider2006gravitational}.

However, conventional EM microlensing has intrinsic limitations because photometric observations primarily constrain the Einstein-radius crossing timescale, which depends on a combination of the lens mass, lens--source geometry, and relative transverse velocity rather than determining these quantities individually~\citep{mao2012astrophysical}.
Additional observables, such as microlensing parallax, finite-source effects, and astrometric deflections, can help break these degeneracies~\citep{gould1992extending,gould2000natural,dominik2000astrometric,an2002first,gould2004resolution}, but are available only for a limited subset of events.
Moreover, photometric microlensing surveys are primarily sensitive to sufficiently bright and well-monitored sources, predominantly accessible in the nearby Universe, limiting the range of sources and lensing configurations that can be probed~\citep{gaudi2012microlensing}.
These limitations motivate complementary probes of compact dark objects.

Gravitational-wave (GW) microlensing provides such a complementary probe.
Compact lenses can imprint wave-optics effects on GW signals, producing frequency-dependent diffraction and interference features~\citep{deguchi1986diffraction,nakamura1998gravitational,takahashi2003wave,PhysRevD.90.062003,PhysRevD.98.083005,PhysRevD.98.103022,PhysRevLett.122.041103}. 
For ground-based GW detectors, these effects are particularly relevant for lens masses in the range $\sim1$--$10^5~M_\odot$, depending on the GW frequency and lens--source geometry.

In this regime of wave optics, the lensing signature is encoded directly in the waveform rather than solely in the magnification of an EM source, providing information that is complementary to conventional photometric microlensing. 
Moreover, GWs can be observed from sources at cosmological distances and are unaffected by dust extinction, allowing compact lenses along distant lines of sight to be probed~\citep{abac2026gwtc_catalog}.

However, individual GW microlensing events provide only limited information, typically constraining only the detector-frame (redshifted) lens mass and leaving a degeneracy between the intrinsic lens mass and lens redshift~\citep{abac2025gwtclensing}.
Recovering the underlying lens population, therefore, requires combining information from multiple detections. 
Hierarchical Bayesian inference offers a framework for doing this by jointly modeling the lens-mass and redshift distributions across an ensemble of events, thereby constraining the hyperparameters that govern the lens population (e.g.,~\cite{gais2022inferring,lin2025constraints}).

To date, no compelling evidence for GW lensing has been reported by the LIGO--Virgo--KAGRA (LVK) Collaboration~\citep{hannuksela2019search,abbott2021search,abbott2024search,abac2025gwtclensing}, although intense scrutiny has been given to the lensed candidate GW231123~\citep{shan2025gw231123,hu2026gw231123,chan2026discovering,chakraborty2026first,goyal2026across,cheung2026diffraction,wang2026gw231123}.
In the absence of confirmed detections, existing studies have instead used GW observations to place constraints on the abundance of compact DM through the non-observation of lensing signatures~\citep{basak2022constraints,barsode2024constraints}.
A complementary question is then \emph{how much can be learned once a population of microlensed GW events becomes available?}
Although intermediate-mass black holes (IMBHs) and globular clusters are unlikely to produce GW microlensing~\citep{Vujeva:2026xfh}, if compact objects are linked to the DM, future GW observatories are expected to detect such events in sufficient numbers, making population-level inference possible~\citep{fairbairn2023microlensing,Vujeva:2025nwg}.

In this Letter, we investigate the extent to which compact-lens populations can be recovered from GW microlensing observations. 
Using astrophysically motivated simulations of microlensed GW events, we perform hierarchical Bayesian inference to jointly constrain the lens-mass distribution and number density while accounting for selection effects.
We then translate these constraints into measurements of the compact-lens abundance, expressed as their contribution to the DM density.

Throughout this work, we adopt the \textsc{Planck15} cosmology~\citep{ade2016planck} as our fiducial model.

\section{Data Setup}~\label{sec:gwmicro}
\subsection{Point-Mass Microlensing}
When a propagating GW signal passes near a compact object with a Schwarzschild radius that is comparable to the GW wavelength, the waveform undergoes frequency-dependent amplification, a phenomenon referred to as \emph{GW microlensing}.
The waveform of a microlensed GW ($h_{\rm ML}$) in the frequency domain can be obtained by multiplying the amplification factor ($F$) by the corresponding unlensed GW waveform ($h_{\rm u}$)~\citep{takahashi2003wave}:
\begin{equation}\label{eq:lensed_wf}
    h_{\rm ML}(w,y) = F(w,y)h_{\rm u}(w),
\end{equation}
where $w \equiv 8\pi GM_{l}^{z} f / c^{3}$ is the dimensionless frequency of the waveform.
$M_{l}^{z} \equiv M_{l}(1+z_{l})$ denotes the redshifted lens mass, with $M_{l}$ and $z_{l}$ the lens-frame mass and redshift, respectively, and $y$ is the dimensionless impact parameter.
These parameters fully specify an isolated point-mass (PM) lens.
The corresponding amplification factor is written analytically as follows~\citep{nakamura1998gravitational,takahashi2003wave}:
\begin{align}
\label{eq:AF_pml}
F(w,y)&=\exp\bigg{\{}\dfrac{\pi w}{4}+i \dfrac{w}{2} \bigg{[}\text{ln} \bigg{(}\dfrac{w}{2}\bigg{)}-2\phi_m(y)\bigg{]}\bigg{\}} \nonumber \\
&\times \Gamma \bigg{(}1- \dfrac{i}{2}w\bigg{)} {_1}F_1\bigg{(}\dfrac{i}{2}w, 1; \dfrac{i}{2}w y^2\bigg{)},
\end{align}
where $_1F_1$ is the confluent hypergeometric function and $\phi_m(y)=(x_m-y)^2/2 - \ln x_m$, where $x_m=(y+\sqrt{y^2+4})/2$.

\subsection{Simulated Microlensed GW Events}
To simulate microlensed GW signals using Eqs.~\eqref{eq:lensed_wf} and \eqref{eq:AF_pml}, we consider only binary black holes (BBHs), whose detection horizon for current ground-based detectors extends to the largest distances among compact-binary sources~\citep{2026arXiv260527226T}.
Assuming a 3-year observing period ($T_{\rm obs}=3,\mathrm{yrs}$), we draw source parameters from the GWTC-3 population models~\citep{abbott2023population}, fixing the hyperparameters to their posterior median values.
Specifically, we adopt the \textsc{PowerLaw+Peak} model for the primary mass distribution and a Madau--Dickinson-like redshift distribution~\citep{madau2014cosmic} over $z_s \in [0.001, 2.3]$, yielding a total of $N_{\rm UL} \simeq 7.9 \times 10^5$ BBHs (see Appendices~\ref{app:component_mass} and~\ref{app:redshift}).
GW signals are then generated using the \texttt{IMRPhenomXPHM-SpinTaylor} waveform model~\citep{pratten2021computationally,PhysRevD.111.104019}.

The resulting synthetic BBH population remains broadly consistent with the population properties inferred from the most recent GWTC-5 catalog, which retains the main features of the BBH mass distribution identified in earlier catalogs~\citep{abbott2023population,abac2025gwtc4pop,2026arXiv260527226T}. 
We further find that our results are largely insensitive to the choice between the GWTC-3 and GWTC-5 population models.
Details of the adopted models and hyperparameters, together with the comparison between the GWTC-3 and GWTC-5 populations, are given in Appendix~\ref{app:gwtc3vs5}.

We draw microlens parameters based on a population described by two hyperparameters: (1) $n_{0}$, the constant comoving number density of microlenses, and (2) $\alpha_{\rm ML}$, the mass-function slope. 
The intrinsic lens mass $M_l$ is distributed as:
\begin{equation}
\label{ppopml}
p(M_{l} \mid \alpha_{\rm ML}) \propto M_{l}^{-\alpha_{\rm ML}},
\end{equation}
motivated by stellar lenses following the initial mass function~\citep{chabrier2003galactic}, of PBHs, for which a power-law mass function arises naturally from scale-invariant density fluctuations or the collapse of cosmic string loops~\citep{carr2017primordial,james2020constraints}, and by compact objects observed by the LVK network~\citep{2026arXiv260527226T}.
We consider lens masses up to $M_l = 10^5~M_{\odot}$, above which lensing effects enter the geometrical-optics regime across the LVK frequency band.

Motivated by~\cite{gais2022inferring}, which showed that an injected IMBH number density of $n_{0}=10^{6}~\mathrm{Mpc}^{-3}$ yields a well-constrained posterior on the number density, we adopt this comoving number density.
For the mass-function slope, we inject $\alpha_{\rm ML} = [1,2]$ to explore two representative population scenarios.

The choice $\alpha_{\rm ML}=1$ corresponds to a log-uniform mass distribution, providing an agnostic baseline model (see, for example, \cite{carr2017primomatterdom} for PBH formation via spectator-field perturbations during an early matter-dominated era).
In contrast, $\alpha_{\rm ML}=2$ describes a steeper distribution in which low-mass lenses dominate the population more strongly, representative of the general class of power-law PBH mass functions~\citep{carr2017primordial,james2020constraints}.
Comparing these two cases allows us to assess how constraints on the underlying lens population depend on $\alpha_{\rm ML}$.

\begin{figure*}[ht]
    \centering
    \includegraphics[width=1.0\linewidth]{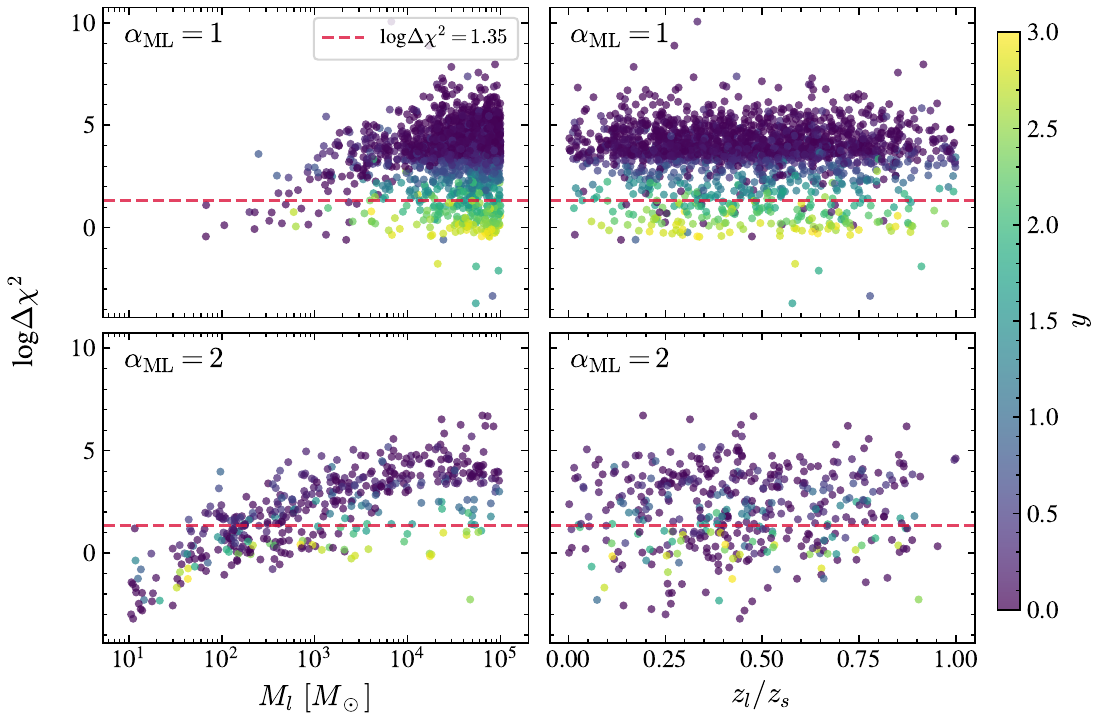}
    \caption{Distributions of $\Delta \chi^{2}$ between microlensed and intrinsic unlensed waveforms, for signals with $\rho_{\rm net}>8$, shown as a function of the intrinsic lens mass ($M_l$) and the ratio of lens to source redshifts ($z_l/z_s$). The color scheme indicates the dimensionless impact parameter ($y$). }\label{fig:MM_dist}
\end{figure*}
Since GW microlensing measurements constrain the detector-frame lens mass, $M_l^z$, we draw $M_l^z$ directly rather than $M_l$.
The microlensing optical depth for a BBH at source redshift $z_s$ is derived in Appendix~\ref{app:tau}.
Combining this optical depth with the assumed $M_l$ distribution in Eq.~\eqref{ppopml}, the probability that a BBH at $z_s$ is microlensed by a lens with $M_l^z$ can be written as:
\begin{equation}
\begin{aligned}
\label{mlz_pop}
    &p(M^{z}_{l} \mid \alpha_{\rm ML}, z_{s}) \\
    & \propto \int^{z_{s}}_{0} dz_{l}\left(M^{z}_{l}\right)^{1-\alpha_{\mathrm{ML}}}(1+z_{l})^{\alpha_{\mathrm{ML}}-2} \frac{ D_{ls}}{D_{s}D_{l}}\frac{dV_{c}}{dz_{l}},
\end{aligned}
\end{equation}
where $V_c$ is the comoving volume, and $D_l$, $D_s$, and $D_{ls}$ are the angular diameter distances from the observer to the lens, from the observer to the source, and from the lens to the source, respectively.

Using Eq.~\eqref{mlz_pop}, we sample $M^{z}_{l}$ within the range $[1,10^{5}]M_{\odot}$.
Note that the source and lens parameters are linked through $z_s$, which determines the lens redshift and $M_l^z$ distribution.
We then sample $y$ from the nearest-effective-lens distribution~\citep{gais2022inferring}, which incorporates the geometric factor ($p(y)\propto y$) arising from the differential lensing area for a spatially uniform source population in the source plane. 
Restricting $y$ to the range [0.01,3.0]\footnote{Extremely small values of $y$ are statistically disfavored because they correspond to a small geometric lensing cross-section, whereas observable microlensing distortions become increasingly suppressed at large $y$.}, we adopt
\begin{equation}
\begin{aligned}
&p\left(y \mid n_0, \alpha_{\rm ML}, z_s\right)\\
&=2 \pi \Sigma\left(n_0, \alpha_{\rm ML}, z_s\right) y e^{-\pi \Sigma\left(n_0, \alpha_{\rm ML}, z_s\right) y^2},
\end{aligned}
\label{eq:y_pop}
\end{equation}
where $\Sigma$ is the effective lens number density within the comoving volume out to $z_s$, projected onto the sky, which is given by
\begin{equation}
\Sigma(n_0,\alpha_{\rm ML},z_s)
=\frac{N_L(z_s)}{4\pi}
\left\langle \theta_E^2 \right\rangle_{\alpha_{\rm ML},z_s},
\end{equation}
where $N_L(z_s) \equiv \int^{z_s}_{0}n_0 dV_c(z_l)$ is the total number of lenses within the comoving volume out to $z_s$, and $\langle\theta_E^2\rangle_{\alpha_{\rm ML},z_s}$ is the mean squared angular Einstein radius averaged over the lens population according to their mass and redshift distributions.

For the $\alpha_{\rm ML}=1$ and $2$ scenarios, the expected numbers of microlensed events given by Eq.~\eqref{eq:total_ML} are $N_{\rm ML}=74{,}933$ and $16{,}275$, respectively. Each microlensed event is then assigned a source parameter set and a corresponding set of lens parameters drawn from the distributions above.

\subsection{Microlensing Detectability}~\label{sec:2.3}
Although the optical depth determines the expected number of microlensed GW events, only a subset of these events produces detectable wave-optics signatures in GW observations.
The detectability of GW microlensing depends not only on the network SNR, $\rho_{\rm net}$, but also on the waveform mismatch $\epsilon$ between microlensed and unlensed signals, which quantifies the degree of waveform distortion induced by wave-optics effects.

Following \cite{ezquiaga2022modified}, we adopt a fiducial distinguishability criterion based on the statistic $\Delta \chi^{2}$, which is interpreted as a measure of how distinguishable a microlensed waveform is from its unlensed counterpart.
In the limit of small waveform mismatches, $\Delta \chi^{2}$ can be approximated as:
\begin{equation}\label{eq:ML_distingusihability}
\Delta \chi^{2}
\simeq
2\rho_{\rm net}^{2} \, \epsilon_{\rm net},
\end{equation}
where $\epsilon_{\rm net}$ denotes the network-averaged mismatch computed from the individual detector signals.
We impose the conventional GW detection threshold, $\rho_{\rm net}>8$, and require an additional distinguishability criterion to identify a detected signal as microlensed. 
Using Eq.~\eqref{eq:ML_distingusihability}, a fiducial mismatch threshold of $\epsilon_{\rm net}>0.03$ at the detection threshold $\rho_{\rm net}=8$ corresponds to $\Delta\chi^{2}>3.84$ (or $\log\Delta\chi^{2}\gtrsim1.35$). 
Signals satisfying both criteria are therefore considered identifiable as microlensed.

Figure~\ref{fig:MM_dist} shows the resulting $\Delta \chi^2$ values between the simulated microlensed waveforms and their corresponding unlensed counterparts for microlensed events that satisfy the SNR detection threshold $\rho_{\rm net}>8$, obtained over three years of simulated observations, as a function of $M_l$ and $z_l/z_s$ for both the $\alpha_{\rm ML}=1$ and $\alpha_{\rm ML}=2$ populations.
In both cases, $\Delta\chi^2$ shows a decreasing trend with increasing $y$, reflecting the progressive suppression of wave-optics distortions at larger $y$.
We estimate $y_{\rm max}$ used in Eq.~\eqref{eq:lensing_optical_depth_micro} based on the adopted $\Delta\chi^2$ threshold.

In addition, a positive correlation between $M_l$ and $\Delta\chi^2$ is evident, as expected from the stronger and more distinguishable waveform distortions produced by more massive lenses.
Since the $\alpha_{\rm ML}=1$ population contains a larger fraction of high $M_l$ lenses than the steeper $\alpha_{\rm ML}=2$ distribution, a correspondingly larger fraction of simulated events satisfy the final distinguishability criterion in the $\alpha_{\rm ML}=1$ case.
The minimum lens mass capable of producing detectable microlensing signatures is $\mathcal{O}(2)~M_\odot$ for $\alpha_{\rm ML}=1$, whereas detectable signals in the $\alpha_{\rm ML}=2$ case can extend down to $\mathcal{O}(1)~M_\odot$.
Moreover, no clear correlation is observed between $\Delta\chi^2$ and $z_l/z_s$. 
While the lensing probability is maximized for lenses located roughly midway between the source and observer due to geometric effects, the strength of observable wave-optics distortions depends more strongly on $M^z_l$ and $y$.

It is important to note that some simulated microlensed signals may satisfy the distinguishability criterion while still failing the network SNR requirement ($\rho_{\rm net} > 8$).
Such events would not be initially detectable.
We therefore retain only events satisfying both detectability and distinguishability criteria and classify them as detectable microlensed GW signals.

\section{Population Inference}\label{sec:method}
\subsection{Selection of Microlensed Events}

From the resulting detectable populations, containing 1727 and 291 events for $\alpha_{\rm ML}=1$ and $2$, respectively, we randomly select 200 events from each population for computationally tractable parameter estimation. 
Random sampling ensures that no additional selection is imposed beyond the detectability criterion. 
We then inject these events into simulated data from the LIGO Livingston and Hanford detectors~\citep{capote2025advanced} and Virgo~\citep{acernese2026advanced}, assuming the expected O4 sensitivities and Gaussian noise realizations~\citep{abbott2020prospects}.

To estimate the parameters of microlensed waveforms, we adopt \texttt{Gravelamps}~\citep{wright2022gravelamps}, which is based on \texttt{bilby}~\citep{ashton2019bilby} and \texttt{dynesty}~\citep{speagle2020dynesty}.
The pipeline yields posterior distributions for $M_l^z$ and $y$, together with the Bayesian evidence for the microlensing hypothesis, $\mathcal{B}_{\rm ML}$.
Performing a parameter estimation for the same signal under the unlensed hypothesis additionally yields the corresponding evidence $\mathcal{B}_{\rm U}$, allowing the Bayes factor between microlensing and unlensed hypotheses, $\mathcal{B}^{\rm ML}_{\rm U}$, to be evaluated.
A microlensed signal is expected to favor the lensing hypothesis with $\mathcal{B}^{\rm ML}_{\rm U} > 1$.
In realistic searches, confident microlensing detections would additionally require a robust false-positive probability (FPP) assessment based on a large population of unlensed background events.
For simplicity, however, we adopt a pragmatic criterion and regard microlensed events with $\mathcal{B}^{\rm ML}_{\rm U}>1$ and well-constrained posteriors for $M_l^z$ and $y$ as confidently identified microlensing events, yielding totals of 158 and 132 events for the $\alpha_{\rm ML}=1$ and $\alpha_{\rm ML}=2$ cases, respectively.
Appendix~\ref{app:mlgw_catalog} compares the lens parameters of identified microlensed events with the underlying population models, illustrating how detectable lens properties depend on the population parameters.
\subsection{Hierarchical Bayesian Framework}
Given the set of detected microlensing events ($\boldsymbol{d}$), the posterior of the hyperparameters ($\lambda$) is obtained through hierarchical Bayesian inference:
\begin{equation}\label{eq:pos_hyperparams}
\begin{aligned}
    &p(\lambda \mid \boldsymbol{d}) \propto p(\lambda)p(N_{\rm{ML},det} \mid \lambda) \\
    & \times \frac{\prod^{N_{\rm{ML},det}}_{i} p(d_{i}|\boldsymbol{\theta_{s,i}},M^{z}_{l,i},y_{i}) p_{\rm pop}(\boldsymbol{\theta_{s,i}},M^{z}_{l,i},y_{i} \mid \lambda)}
     {\beta(\lambda)^{N_{\rm{ML},det}}} ,
\end{aligned}
\end{equation}
where $N_{\rm ML,det}$ is the number of detected microlensed events and $\beta (\lambda)$ indicates the detection efficiency for a given hyperparameter set, corresponding to a selection-bias correction~\citep{mandel2019extracting,farr2019accuracy}.
The hyperparameter set consists of source-population, cosmological, and lens-population parameters, denoted by $\lambda=({\lambda_s, \lambda_c,\lambda_l})$.
Here, $\lambda_s$ describes the BBH population as introduced in Eqs.~\eqref{eq:m1_pop}--\eqref{eq:z_pop}, while $\lambda_c$ denotes the cosmological parameters, such as the Hubble constant $H_0$ and the density parameters $\Omega_m$ and $\Omega_\Lambda$. In this work, the lens-population parameters are $\lambda_l=(\alpha_{\rm ML},n_0)$.

The first term in Eq.~\eqref{eq:pos_hyperparams} is the prior on the hyperparameters.
In this work, we fix $\lambda_s$ and $\lambda_c$ to equal their injected values to focus exclusively on the inference of  $\lambda_l$.
We adopt a broad uniform prior for the lens mass-function slope, $\alpha_{\rm ML} \in \mathrm{U}[-4,12]$. For the lens number density, we adopt $\log_{10} n_{0} \in \mathrm{U}[0,10]$, following the prior range adopted in~\cite{gais2022inferring}.

\begin{figure*}[ht]
    \centering
    \includegraphics[width=0.495\linewidth]{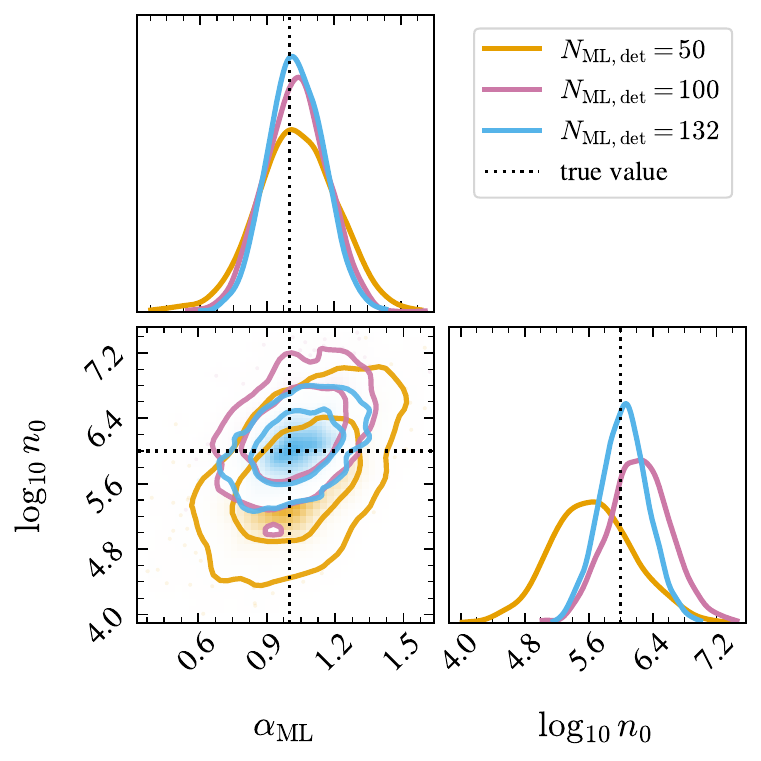}
    \includegraphics[width=0.495\linewidth]{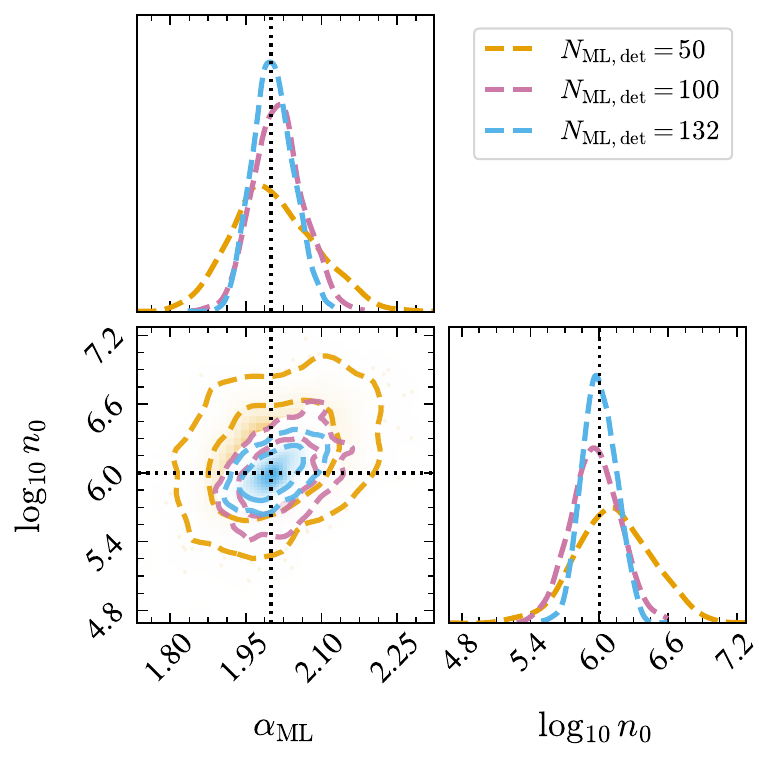}
    \caption{Corner plots showing the posteriors of $\alpha_{\rm ML}$ and $\log_{10}n_0$ inferred from the $N_{\rm ML,det}=50$, $100$, and $132$ cases for the two fiducial scenarios $\alpha_{\rm ML}=1$ (left) and $\alpha_{\rm ML}=2$ (right).
    The black dotted lines denote the corresponding true values.}\label{fig:alphaML_n0}
\end{figure*}
The second term describes the probability of detecting $N_{\rm ML,det}$ microlensing events for a given set of $\lambda$, assuming a Poisson distribution with an expected number given by:
\begin{equation}
\begin{aligned}
    &\langle N_{\rm ML, det}(\lambda) \rangle =N_{\rm ML}(\lambda) \\ & \times \int p_{\rm pop}(\boldsymbol{\theta_{s}},M^{z}_{l},y \mid \lambda) P_{\rm det}(\boldsymbol{\theta_{s}},M^{z}_{l},y) ~d\boldsymbol{\theta_{s}}dM^{z}_{l}dy
\end{aligned}
 \end{equation}
where $N_{\rm ML}(\lambda)$ is the total number of microlensed events defined in Eq.~\eqref{eq:total_ML}, while $p_{\rm pop}(\boldsymbol{\theta_s}, M_l^z, y\mid\lambda)$ encodes the underlying astrophysical population model for the source and lens parameters.
In this work, the lens population is described by $M^z_l$ distribution given in Eq.~\eqref{mlz_pop}, together with $y$ distribution given in Eq.~\eqref{eq:y_pop}.
The term $P_{\rm det}$ denotes the probability that a microlensed signal with the given source and lens parameters satisfies the adopted detection and distinguishability criteria.

The third term in Eq.~\eqref{eq:pos_hyperparams} represents the joint likelihood of all detected microlensed events, constructed as a product of individual event contributions, assuming that each event is statistically independent given $\lambda$.
This assumption allows the joint likelihood to be factorized into a product of event-level terms.
In the numerator, $p(d_i\mid \boldsymbol{\theta_{s,i}}, M_{l,i}^z, y_i)$ denotes the event-level likelihood of each set of observed microlensed GW data, evaluated using posterior samples obtained from parameter estimation.
In the denominator, $\beta(\lambda)$ denotes the detection efficiency, defined as the probability that a source drawn from the underlying population is identified as a microlensed event satisfying both the adopted SNR detection threshold and distinguishability criterion. 
It therefore accounts for the selection effects associated with both requirements.
The detection efficiency for microlensed signals can be written as:
\begin{equation}
\begin{aligned}
\label{eq:lens_selection_eff}
    \beta(\lambda) &\equiv \frac{\langle N_{\rm ML,det}\left(\lambda\right) \rangle}{N_{\rm ML}}\\
    &=  \int p_{\rm pop}(\boldsymbol{\theta_{s}},M^{z}_{l},y \mid \lambda) P_{\rm det}(\boldsymbol{\theta_{s}},M^{z}_{l},y).
\end{aligned}
\end{equation}

In practice, Eq.~\eqref{eq:lens_selection_eff} can be estimated using importance sampling over a large set of simulated microlensed injections:
\begin{equation}
\beta(\lambda)
\simeq
\frac{1}{N_{\rm inj}}
\sum_{i}^{N_{\rm inj,det}}
\frac{
p_{\rm pop}(\boldsymbol{\theta_{s,i}},M^{z}_{l,i},y_{i}\mid \lambda)
}
{\pi(\boldsymbol{\theta_{s,i}},M^{z}_{l,i},y_{i})},
\end{equation}
where $N_{\rm inj}$ and $N_{\rm inj,det}$ denote the total number of injected and successfully detected microlensed signals, respectively, and $\pi$ is the joint prior distribution used in the injection campaign.
\section{Results}
\subsection{Inference of the Compact-Lens Distribution}\label{sec:4.1}

Using Eq.~\eqref{eq:pos_hyperparams}, we infer the joint posterior distributions for $\alpha_{\rm ML}$ and $n_0$.
To examine how the constraints on these population parameters improve as the mock microlensed-event sample accumulates, we consider cumulative samples containing $N_{\rm ML,det}=50$, $100$, and $132$ events for both the $\alpha_{\rm ML}=1$ and $2$ scenarios.
The $\alpha_{\rm ML}=2$ scenario has a lower detectable-event rate than the $\alpha_{\rm ML}=1$ scenario, and therefore requires a longer observing time to accumulate the same number of events.
We compare the two scenarios at fixed $N_{\rm ML,det}$ to control for differences in sample size and isolate the impact of the underlying lens-mass distribution in Eq.~\eqref{ppopml} on the population constraints.

As shown in Figure~\ref{fig:alphaML_n0}, the widths of the posteriors for both hyperparameters become progressively narrower as the number of detections increases, while the posterior peaks converge toward their respective true values.
Notably, $n_0$ is less well constrained than $\alpha_{\rm ML}$. 
This difference arises from its strong degeneracy with $y$, which more directly governs the microlensing probability and magnification than $M_l^z$.

Furthermore, across both the $\alpha_{\rm ML}=1$ and $\alpha_{\rm ML}=2$ scenarios, the posteriors show a positive correlation between $\alpha_{\rm ML}$ and $n_0$, regardless of $N_{\rm ML,det}$.
Physically, a larger $\alpha_{\rm ML}$ shifts the lens population toward lower-mass compact objects, which generally produce weaker microlensing signatures and are therefore less likely to be detected.
A higher $n_0$ is then required to reproduce the same number of detections, giving rise to the observed $\alpha_{\rm ML}$--$n_0$ degeneracy.
Increasing the catalog size tightens the constraints on both parameters, but the correlation persists, since it arises from the structure of the detection rate rather than from limited statistics alone.

In the $\alpha_{\rm ML}=2$ scenario, the intrinsic lens-mass distribution is strongly weighted towards the lower-mass end.
Even for $N_{\rm ML,det}=50$, the detected population is therefore dominated by low-mass lenses, and the absence of high-mass lenses provides a strong constraint on $\alpha_{\rm ML}$.
Because of the $\alpha_{\rm ML}$--$n_0$ correlation, this in turn tightens the constraint on $n_0$.

For each dataset simulated under $\alpha_{\rm ML}=1$ or
$\alpha_{\rm ML}=2$, we calculate the Bayes factor between the
two hypotheses by analyzing them separately with
$\alpha_{\rm ML}$ fixed to each value. 
We denote the log Bayes
factor in favor of the $\alpha_{\rm ML}=1$ and
$\alpha_{\rm ML}=2$ hypotheses by $\ln\mathcal{B}^{1}_{2}$
and $\ln\mathcal{B}^{2}_{1}$, respectively. 
The resulting Bayes factors are reported in Table~\ref{tab:bf_alpha}.
As expected, the Bayes factor in favor of the injected
hypothesis increases with $N_{\rm ML,det}$.

\begin{table}[t]
    \centering
    \begin{tabular}{lccc}
        \toprule
        & \multicolumn{3}{c}{$N_{\rm ML,det}$} \\
        \cmidrule(lr){2-4}
        Injected $\alpha_{\rm ML}$ (Bayes factor)
        & $50$ & $100$ & $132$ \\
        \midrule
        $1$ ($\ln{\mathcal{B}^{1}_{2}}$) & $69$ & $127$ & $161$ \\
        $2$ ($\ln{\mathcal{B}^{2}_{1}}$) & $76$ & $159$ & $226$ \\
        \bottomrule
    \end{tabular}
    \caption{Log Bayes factors in favor of the injected
    $\alpha_{\rm ML}$ hypothesis for different $N_{\rm ML,det}$.
    Each entry is defined as
    $\ln\mathcal{B}^{\rm inj}_{\rm alt}=\ln Z_{\rm inj}-\ln Z_{\rm alt}$,
    where $Z_{\rm inj}$ and $Z_{\rm alt}$ are the Bayesian
    evidences for the injected and alternative fixed-$\alpha_{\rm ML}$
    hypotheses, respectively.}\label{tab:bf_alpha}
\end{table}
\subsection{Constraining the Dark Matter Fraction in Microlenses}\label{sec:4.2}

\begin{figure*}[t]
    \centering
    \includegraphics[width=1.0\linewidth]{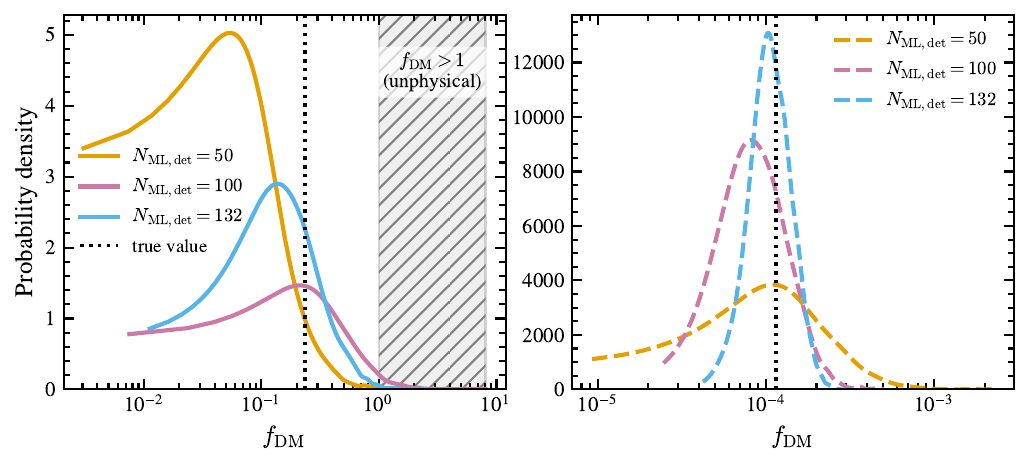}
    \caption{Posteriors of $f_{\rm DM}$ inferred for the $\alpha_{\rm ML}=1$ (left) and $\alpha_{\rm ML}=2$ (right) scenarios using catalogs with $N_{\rm ML,det}=50$, $100$, and $132$ detected microlensing events.
    The hatched gray region marks the unphysical range with $f_{\rm DM}>1$.
    The black dotted lines denote the corresponding true values.
    }\label{fig:fdm_all}
\end{figure*}
According to~\cite{turner1984statistics}, the comoving number density $n_0$ appearing in Eq.~\eqref{eq:lensing_optical_depth_micro} can be written as:
\begin{equation}\label{number_density}
\begin{aligned}
    n_0 &= \frac{\rho_{l}}{\langle M_l \rangle}\\ 
    &=\frac{\rho_c \Omega_{l}}{\langle M_l \rangle} \\
    &=\frac{3H^2_0 \Omega_l}{8\pi G\langle M_l \rangle},
\end{aligned}
\end{equation}
where $\rho_l$ and $\rho_c$ denote the present-day compact-lens density and the critical density of the Universe, respectively, and $\Omega_l \equiv \rho_l/\rho_c$ is the corresponding density parameter.
$\langle M_l \rangle$ denotes the expectation value of the intrinsic lens mass, analytically derived from the assumed mass distribution $p(M_l \mid \alpha_{\rm ML})$ truncated to a specific mass range.
For a given $\alpha_{\rm ML}$, the theoretical $\langle M_l \rangle$ is given by:
\begin{equation}
\label{eq:mean_ml}
\langle M_l \rangle
=
\begin{cases}
\displaystyle
\left(
\frac{1-\alpha_{\rm ML}}
{2-\alpha_{\rm ML}}
\right)
\frac{
M_{\rm max}^{\,2-\alpha_{\rm ML}}
-
M_{\rm min}^{\,2-\alpha_{\rm ML}}
}{
M_{\rm max}^{\,1-\alpha_{\rm ML}}
-
M_{\rm min}^{\,1-\alpha_{\rm ML}}
},
&
\alpha_{\rm ML} \neq 1,\,2,
\\[14pt]
\displaystyle
\frac{
M_{\rm max}-M_{\rm min}
}{
\ln(M_{\rm max}/M_{\rm min})
},
&
\alpha_{\rm ML} = 1,
\\[14pt]
\displaystyle
\left(
\frac{
M_{\rm min}M_{\rm max}
}{
M_{\rm max}-M_{\rm min}
}
\right)
\ln\left(
\frac{M_{\rm max}}{M_{\rm min}}
\right),
&
\alpha_{\rm ML} = 2.
\end{cases}
\end{equation}
where $M_{\rm min}$ and $M_{\rm max}$ are the lower and upper bounds
of the intrinsic lens-mass distribution, respectively. Since we consider
redshifted lens masses $M_l^z \in [1,10^5]\,M_\odot$ and the maximum
source redshift is $z_{s,\rm max}=2.3$, we adopt the corresponding
global bounds on the intrinsic lens mass,
$M_{\rm min}=1/(1+z_{s,\rm max})\,M_\odot \simeq 0.303\,M_\odot$
and $M_{\rm max}=10^5\,M_\odot$.
If the compact lenses are interpreted as DM objects, such as PBHs or DM clumps, their density parameter can be expressed as $\Omega_l \equiv f_{\rm DM}\Omega_{\rm DM}$, where $f_{\rm DM}$ denotes the fraction of total DM density contributed by compact
lenses within the mass range [$M_{\rm min},~M_{\rm max}$].
Note that $f_{\rm DM}=1$ corresponds to the case in which all DM is contained in compact lenses within this mass range.

Consequently, the posterior constraints on $n_0$ and $\alpha_{\rm ML}$ obtained from the hierarchical Bayesian analysis in Section~\ref{sec:4.1} can be translated into constraints on $f_{\rm DM}$, assuming a fiducial DM density parameter $\Omega_{\rm DM} \simeq 0.26$~\citep{ade2016planck}.
Under these assumptions, the DM fraction associated with compact lenses can be written as:
\begin{equation}\label{eq:fdm}
    f_{\rm DM} = \frac{8\pi G n_0 \langle M_l \rangle}{3 H_0^2 \Omega_{\rm DM}}.
\end{equation}
Unlike monochromatic-mass constraints, our analysis accounts for an extended lens-mass range and its power-law spectrum, allowing the inferred $f_{\rm DM}$ to reflect both the lens abundance and its mass distribution.

Figure~\ref{fig:fdm_all} shows the posteriors of $f_{\rm DM}$, derived from the posterior samples of $n_0$ and $\alpha_{\rm ML}$ for each $N_{\rm ML,det}$ case.
In the $\alpha_{\rm ML}=1$ scenario, the relatively high mean lens mass leads to a comparatively large true DM fraction of $f_{\rm DM}\simeq0.23$.
For $N_{\rm ML,det}=50$, the posterior is weakly constrained, with substantial support over a broad range and its peak noticeably offset from the true value. 
As shown similarly in Figure~\ref{fig:alphaML_n0}, the posterior becomes progressively narrower and more closely centered on the true value as $N_{\rm ML,det}$ increases.
In particular, the lower bound on $f_{\rm DM}$ becomes progressively more stringent with increasing $N_{\rm ML,det}$.

In contrast, for the $\alpha_{\rm ML}=2$ scenario, the substantially lower mean lens mass yields a much smaller true DM fraction of $f_{\rm DM}\simeq1.2\times10^{-4}$. 
The posterior is already peaked near the true value for $N_{\rm ML,det}=50$, reflecting the relatively strong constraint on $\alpha_{\rm ML}$. 

It is worth noting that while the physical boundary strictly requires $f_{\rm DM} \leq 1$, our posterior samples for the $\alpha_{\rm ML}=1$ scenario extend into the unphysical regime ($f_{\rm DM}>1$).
This issue arises because the analysis was performed using broad hyperparameter priors that do not explicitly impose the physical constraint on $f_{\rm DM}$.
In our simplified population model, the physically allowed region in the $(n_0,\alpha_{\rm ML})$ parameter space can be derived analytically from Eq.~\eqref{eq:fdm}. 
In more realistic compact-object population models, however, the mapping between hyperparameters and the implied value of $f_{\rm DM}$ is expected to become substantially more complex. 
For this reason, adopting broad priors at the hyperparameter level and subsequently excluding posterior samples with $f_{\rm DM}>1$ provides a more flexible and model-independent approach than imposing hard physical boundaries a priori.

In addition to $f_{\rm DM}$, which quantifies the total contribution of compact lenses across the given lens-mass range, the differential dark-matter fraction as a function of lens mass, $d f_{\rm DM}/dM_l$, provides a mass-resolved view of this contribution.
In our power-law lens mass distribution model, this quantity can be derived analytically. 
In particular,
\begin{equation}
\frac{d f_{\rm DM}}{d\log_{10} M_l}
\propto M_l^{2-\alpha_{\rm ML}},
\end{equation}
where $\alpha_{\rm ML}=2$ corresponds to an equal contribution to the DM density from each logarithmic mass interval.
For $\alpha_{\rm ML}<2$, the DM contribution is weighted towards higher lens masses, whereas $\alpha_{\rm ML}>2$ shifts the contribution towards lower masses.
Figure~\ref{fig:fdm_diff} in Appendix~\ref{app:diff_fdm} illustrates this dependence.
\section{Summary and Discussion}
We have investigated the feasibility of constraining compact-object populations through hierarchical Bayesian inference applied to a catalog of identified GW microlensing events, while consistently accounting for selection effects.
Using simulated microlensed BBH signals drawn from two representative power-law lens-mass distributions, with $\alpha_{\rm ML}=1$ and $\alpha_{\rm ML}=2$, we demonstrate that the accumulation of microlensed GW detections can place meaningful constraints on both the mass-function slope $\alpha_{\rm ML}$ and the comoving lens number density $n_0$, while simultaneously breaking the degeneracy between these hyperparameters.

We further translate the inferred population parameters into constraints on the compact-DM fraction, $f_{\rm DM}$, and investigate its dependence on both microlens mass and redshift.
For a fixed number of detections, steeper microlens mass functions yield tighter constraints on $f_{\rm DM}$, implying that compact-object populations dominated by lower-mass lenses are intrinsically more informative for measuring it.
Accordingly, both the differential compact-DM fraction and the redshift-binned compact-DM fraction are recovered with smaller uncertainties for the $\alpha_{\rm ML}= 2$ scenario than for the $\alpha_{\rm ML}= 1$ scenario.

An additional advantage of our hierarchical framework is that it does not require explicit hard priors derived from the physical bounds on $f_{\rm DM}$.
Constructing such priors becomes increasingly difficult as the relationship between the population parameters and the physically allowed $f_{\rm DM}$ grows more complicated in realistic population models.
Instead, the physically allowed and disallowed regions can be identified directly from the inferred $f_{\rm DM}$ posterior, making the framework both flexible and readily extensible to more general compact-object population models.
Moreover, the inferred $f_{\rm DM}$ as a function of lens mass and redshift can reveal how the abundance of compact objects in different mass ranges evolves across cosmic time.
Therefore, if observations can probe sufficiently distant lenses, the redshift evolution of $f_{\rm DM}(z_l)$ beyond the peak of the cosmic star-formation rate can provide a test of whether the inferred compact-object population is consistent with conventional stellar remnants alone or requires additional populations such as PBHs.

Although our framework provides a robust foundation for population inference, several simplifying assumptions adopted in this work motivate important directions for future refinement.
We employed a simplified population configuration in which the source-population hyperparameters are fixed, thereby focusing exclusively on the lens population.
Incorporating uncertainties in the source population would weaken the constraints on the population properties. 
However, such an extension would also enable a direct investigation of potential degeneracies between the source and lens populations, particularly in scenarios where both originate from dark compact objects.
This possibility becomes especially relevant if a fraction of the BBH population detected by LVK is itself primordial in origin~\citep{chen2022confronting,liu2023constraining}, since the source and lens populations could then share common formation channels and population characteristics.

In addition, we adopted a simple power-law model for the intrinsic lens-mass distribution. 
More physically motivated population distributions should be considered for PBH and other DM-related object populations when their theoretical predictions are sufficiently well established (see e.g., \cite{PhysRevD.94.063530,fairbairn2023microlensing,bagui2025primordial,Vujeva:2025nwg,ho2026diffraction}).
We also assumed a constant $n_0$, despite the possibility that the compact-lens population evolves over cosmic time.
A more realistic treatment could incorporate redshift evolution through a parameterization, such as $n_l(z)=n_0(1+z)^{\kappa}$, where $\kappa$ characterizes the redshift evolution of the lens abundance.
Such models generally require additional hyperparameters and would introduce further correlations between the source and lens population properties.

Regarding detectability, two aspects of the selection effects should be further refined.
First, current GW search pipelines may fail to detect highly distorted microlensed signals, even when their optimal SNRs are sufficiently large, because signal-consistency tests can downrank waveforms that deviate substantially from the unlensed template bank~\citep{chan2024detectability}.
A realistic selection function should therefore account for the additional detection criteria used by GW search pipelines, such as the $\xi^2$ signal-consistency test~\citep{babak2013searching}. 
Second, among the events recovered by the search pipeline, identifying a signal as microlensed requires a sufficiently low FPP.
Rather than relying solely on a Bayes-factor threshold, this probability should be calibrated through dedicated background injections, allowing the microlensing identification criterion and its associated selection effects to be modeled consistently.

Lastly, our injection-and-recovery framework relies exclusively on the isolated point-mass lens model.
However, microlensed GW signals may also originate from galaxies in which strong lensing by their gravitational potential is possible, with stellar or compact-object populations embedded within the lens galaxy introducing additional microlensing perturbations.
Such systems can produce waveform distortions substantially more complex than those generated by an isolated point-mass lens alone~\citep{diego2019observational,PhysRevD.101.123512,cheung2021stellar,mishra2021gravitational,meena2022gravitational,seo2022improving,yeung2023detectability,shan2023microlensing,Shan:2023qvd,seo2025residual,goyal2026across,Vujeva:2026xfh}. 
Similar diffraction is also expected near caustics~\citep{Ezquiaga:2025gkd}, the regions of highest magnification, leading to unique observational signatures~\citep{Lo:2024wqm} and boosting the effects of DM subhalos~\citep{Vujeva:2025nwg}.
Addressing these limitations will require both more robust statistical detection frameworks and the development of more sophisticated waveform models in future analyses.

\begin{acknowledgments}
We thank Srashti Goyal for her valuable comments and feedback.
E. S. is supported by grants from the College of Science and Engineering of the University of Glasgow.
J.M.E. is supported by Research Grant No.~VIL53101 from Villum Fonden and the Marie Sklodowska-Curie Grant Agreement No.~847523 INTERACTIONS.
The Center of Gravity is a Center of Excellence funded by the Danish National Research Foundation under grant no.~DNRF184..
M. H. is supported by the Science and Technology Facilities Council (Grant Ref. ST/V005634/1).
The authors are grateful for computational resources provided by the LIGO Laboratory and supported by the National Science Foundation Grants PHY-0757058 and PHY-0823459.
This material is based upon work supported by NSF's LIGO Laboratory which is a major facility fully funded by the National Science Foundation.
\end{acknowledgments}

\bibliography{main}{}

@article{gais2022inferring,
  title={Inferring the intermediate-mass black hole number density from gravitational-wave lensing statistics},
  author={Gais, Joseph and Ng, Ken KY and Seo, Eungwang and Wong, Kaze WK and Li, Tjonnie GF},
  journal={The Astrophysical Journal Letters},
  volume={932},
  number={1},
  pages={L4},
  year={2022},
  publisher={The American Astronomical Society}
}

@article{Vujeva:2026xfh,
    author = "Vujeva, Luka and Ezquiaga, Jose Mar{\'\i}a and Lo, Rico K. L. and Zwick, Lorenz",
    title = "{Probability of gravitational-wave lensing by intermediate-mass black holes and globular clusters}",
    eprint = "2608.07667",
    archivePrefix = "arXiv",
    primaryClass = "astro-ph.CO",
    month = "8",
    year = "2026"
}

@article{Vujeva:2025nwg,
    author = "Vujeva, Luka and Ezquiaga, Jose Mar{\'\i}a and Gilman, Daniel and Goyal, Srashti and Zumalac{\'a}rregui, Miguel",
    title = "{Dark matter subhalos and higher order catastrophes in gravitational wave lensing}",
    eprint = "2510.14953",
    archivePrefix = "arXiv",
    primaryClass = "astro-ph.CO",
    doi = "10.1103/kp86-q1h2",
    journal = "Phys. Rev. D",
    volume = "114",
    number = "8",
    pages = "083024",
    year = "2026"
}

@article{Ezquiaga:2025gkd,
    author = "Ezquiaga, Jose Mar{\'\i}a and Lo, Rico K. L. and Vujeva, Luka",
    title = "{Diffraction around caustics in gravitational wave lensing}",
    eprint = "2503.22648",
    archivePrefix = "arXiv",
    primaryClass = "gr-qc",
    doi = "10.1103/tkk2-x9st",
    journal = "Phys. Rev. D",
    volume = "112",
    number = "4",
    pages = "043544",
    year = "2025"
}

@article{Lo:2024wqm,
    author = "Lo, Rico K. L. and Vujeva, Luka and Ezquiaga, Jose Mar{\'\i}a and Chan, Juno C. L.",
    title = "{Observational Signatures of Highly Magnified Gravitational Waves from Compact Binary Coalescence}",
    eprint = "2407.17547",
    archivePrefix = "arXiv",
    primaryClass = "gr-qc",
    doi = "10.1103/PhysRevLett.134.151401",
    journal = "Phys. Rev. Lett.",
    volume = "134",
    number = "15",
    pages = "151401",
    year = "2025"
}

@article{Caliskan:2023zqm,
    author = "{\c{C}}al{\i}{\c{s}}kan, Mesut and Anil Kumar, Neha and Ji, Lingyuan and Ezquiaga, Jose M. and Cotesta, Roberto and Berti, Emanuele and Kamionkowski, Marc",
    title = "{Probing wave-optics effects and low-mass dark matter halos with lensing of gravitational waves from massive black holes}",
    eprint = "2307.06990",
    archivePrefix = "arXiv",
    primaryClass = "astro-ph.CO",
    doi = "10.1103/PhysRevD.108.123543",
    journal = "Phys. Rev. D",
    volume = "108",
    number = "12",
    pages = "123543",
    year = "2023"
}

@article{pratten2021computationally,
  title={Computationally efficient models for the dominant and subdominant harmonic modes of precessing binary black holes},
  author={Pratten, Geraint and Garc{\'\i}a-Quir{\'o}s, Cecilio and Colleoni, Marta and Ramos-Buades, Antoni and Estell{\'e}s, H{\'e}ctor and Mateu-Lucena, Maite and Jaume, Rafel and Haney, Maria and Keitel, David and Thompson, Jonathan E and others},
  journal={Physical Review D},
  volume={103},
  number={10},
  pages={104056},
  year={2021},
  publisher={APS}
}

@article{PhysRevD.111.104019,
  title = {Fast frequency-domain gravitational waveforms for precessing binaries with a new twist},
  author = {Colleoni, Marta and Ramis Vidal, Felip A. and Garc\'{\i}a-Quir\'os, Cecilio and Ak\ifmmode \mbox{\c{c}}\else \c{c}\fi{}ay, Sarp and Bera, Sayantani},
  journal = {Phys. Rev. D},
  volume = {111},
  issue = {10},
  pages = {104019},
  numpages = {17},
  year = {2025},
  month = {May},
  publisher = {American Physical Society},
  doi = {10.1103/PhysRevD.111.104019},
  url = {https://link.aps.org/doi/10.1103/PhysRevD.111.104019}
}

@article{ezquiaga2022modified,
  title={Modified gravitational wave propagation with higher modes and its degeneracies with lensing},
  author={Ezquiaga, Jose Maria and Hu, Wayne and Lagos, Macarena and Lin, Meng-Xiang and Xu, Fei},
  journal={Journal of Cosmology and Astroparticle Physics},
  volume={2022},
  number={08},
  pages={016},
  year={2022},
  publisher={IOP Publishing}
}

@article{turner1984statistics,
  title={The statistics of gravitational lenses-The distributions of image angular separations and lens redshifts},
  author={Turner, Edwin L and Ostriker, Jeremiah P and Gott III, J Richard},
  journal={Astrophysical Journal, Part 1 (ISSN 0004-637X), vol. 284, Sept. 1, 1984, p. 1-22.},
  volume={284},
  pages={1--22},
  year={1984}
}

@article{ashton2019bilby,
  title={BILBY: A user-friendly Bayesian inference library for gravitational-wave astronomy},
  author={Ashton, Gregory and H{\"u}bner, Moritz and Lasky, Paul D and Talbot, Colm and Ackley, Kendall and Biscoveanu, Sylvia and Chu, Qi and Divakarla, Atul and Easter, Paul J and Goncharov, Boris and others},
  journal={The Astrophysical Journal Supplement Series},
  volume={241},
  number={2},
  pages={27},
  year={2019},
  publisher={The American Astronomical Society}
}

@article{wright2022gravelamps,
  title={Gravelamps: gravitational wave lensing mass profile model selection},
  author={Wright, Mick and Hendry, Martin},
  journal={The Astrophysical Journal},
  volume={935},
  number={2},
  pages={68},
  year={2022},
  publisher={The American Astronomical Society}
}

@article{barsode2024constraints,
  title={Constraints on compact dark matter from the nonobservation of gravitational-wave strong lensing},
  author={Barsode, A and Kapadia, SJ and Ajith, Parameswaran},
  journal={The Astrophysical Journal},
  volume={975},
  number={1},
  pages={48},
  year={2024},
  publisher={The American Astronomical Society}
}

@article{sasaki2018primordial,
  title={Primordial black holes—perspectives in gravitational wave astronomy},
  author={Sasaki, Misao and Suyama, Teruaki and Tanaka, Takahiro and Yokoyama, Shuichiro},
  journal={Classical and Quantum Gravity},
  volume={35},
  number={6},
  pages={063001},
  year={2018},
  publisher={IOP Publishing}
}

@article{fairbairn2023microlensing,
  title={Microlensing of gravitational waves by dark matter structures},
  author={Fairbairn, Malcolm and Urrutia, Juan and Vaskonen, Ville},
  journal={Journal of Cosmology and Astroparticle Physics},
  volume={2023},
  number={07},
  pages={007},
  year={2023},
  publisher={IOP Publishing}
}

@article{basak2022constraints,
  title={Constraints on compact dark matter from gravitational wave microlensing},
  author={Basak, Soummyadip and Ganguly, A and Haris, K and Kapadia, S and Mehta, AK and Ajith, P},
  journal={The Astrophysical Journal Letters},
  volume={926},
  number={2},
  pages={L28},
  year={2022},
  publisher={The American Astronomical Society}
}

@article{ade2016planck,
  title={Planck 2015 results-xiii. cosmological parameters},
  author={Ade, Peter AR and Aghanim, N and Arnaud, M and Ashdown, Mark and Aumont, J and Baccigalupi, C and Banday, AJ and Barreiro, RB and Bartlett, JG and Bartolo, N and others},
  journal={Astronomy \& Astrophysics},
  volume={594},
  pages={A13},
  year={2016},
  publisher={EDP sciences}
}

@article{abac2025gwtclensing,
  title={GWTC-4.0: Searches for Gravitational-Wave Lensing Signatures},
  author={Abac, AG and Abouelfettouh, I and Acernese, F and Ackley, K and Adamcewicz, C and Adhicary, S and Adhikari, D and Adhikari, N and Adhikari, RX and Adkins, VK and others},
  journal={arXiv preprint arXiv:2512.16347},
  year={2025}
}

@article{lin2025constraints,
  title={Constraints on Compact Dark Matter Population from Micro-lensing Effect of Gravitational Wave for the third-generation gravitational Wave Detector},
  author={Lin, Xin-Yi and Wang, Xi-Jing and Zhou, Huan and Li, Zhengxiang and Liao, Kai and Zhu, Zong-Hong},
  journal={arXiv preprint arXiv:2508.13577},
  year={2025}
}

@article{takahashi2003wave,
  title={Wave effects in the gravitational lensing of gravitational waves from chirping binaries},
  author={Takahashi, Ryuichi and Nakamura, Takashi},
  journal={The Astrophysical Journal},
  volume={595},
  number={2},
  pages={1039},
  year={2003},
  publisher={IOP Publishing}
}

@article{abbott2023population,
  title={Population of merging compact binaries inferred using gravitational waves through GWTC-3},
  author={Abbott, Richard and Abbott, TD and Acernese, F and Ackley, K and Adams, C and Adhikari, N and Adhikari, RX and Adya, VB and Affeldt, C and Agarwal, D and others},
  journal={Physical Review X},
  volume={13},
  number={1},
  pages={011048},
  year={2023},
  publisher={APS}
}

@article{chabrier2003galactic,
  title={Galactic stellar and substellar initial mass function},
  author={Chabrier, Gilles},
  journal={Publications of the Astronomical Society of the Pacific},
  volume={115},
  number={809},
  pages={763--795},
  year={2003},
  publisher={The University of Chicago Press}
}

@article{abac2025gwtc4pop,
  title={GWTC-4.0: Population properties of merging compact binaries},
  author={Abac, AG and Abouelfettouh, I and Acernese, F and Ackley, K and Adamcewicz, C and Adhicary, S and Adhikari, D and Adhikari, N and Adhikari, RX and Adkins, VK and others},
  journal={arXiv preprint arXiv:2508.18083},
  year={2025}
}

@article{abbott2020prospects,
  title={Prospects for observing and localizing gravitational-wave transients with Advanced LIGO, Advanced Virgo and KAGRA},
  author={Abbott, Benjamin P and Abbott, R and Abbott, TD and Abraham, S and Acernese, F and Ackley, K and Adams, C and Adya, VB and Affeldt, C and Agathos, M and others},
  journal={Living reviews in relativity},
  volume={23},
  pages={1--69},
  year={2020},
  publisher={Springer}
}

@article{deguchi1986diffraction,
  title={Diffraction in gravitational lensing for compact objects of low mass},
  author={Deguchi, Shuji and Watson, William D},
  journal={Astrophysical Journal, Part 1 (ISSN 0004-637X), vol. 307, Aug. 1, 1986, p. 30-37.},
  volume={307},
  pages={30--37},
  year={1986}
}

@article{nakamura1998gravitational,
  title={Gravitational lensing of gravitational waves from inspiraling binaries by a point mass lens},
  author={Nakamura, Takahiro T},
  journal={Physical review letters},
  volume={80},
  number={6},
  pages={1138},
  year={1998},
  publisher={APS}
}

@article{PhysRevD.90.062003,
  title = {Gravitational lensing effects on parameter estimation in gravitational wave detection with advanced detectors},
  author = {Cao, Zhoujian and Li, Li-Fang and Wang, Yan},
  journal = {Phys. Rev. D},
  volume = {90},
  issue = {6},
  pages = {062003},
  numpages = {12},
  year = {2014},
  month = {Sep},
  publisher = {American Physical Society},
  doi = {10.1103/PhysRevD.90.062003},
  url = {https://link.aps.org/doi/10.1103/PhysRevD.90.062003}
}

@article{PhysRevLett.122.041103,
  title = {Gravitational-Wave Fringes at LIGO: Detecting Compact Dark Matter by Gravitational Lensing},
  author = {Jung, Sunghoon and Shin, Chang Sub},
  journal = {Phys. Rev. Lett.},
  volume = {122},
  issue = {4},
  pages = {041103},
  numpages = {6},
  year = {2019},
  month = {Jan},
  publisher = {American Physical Society},
  doi = {10.1103/PhysRevLett.122.041103},
  url = {https://link.aps.org/doi/10.1103/PhysRevLett.122.041103}
}

@article{speagle2020dynesty,
  title={dynesty: a dynamic nested sampling package for estimating Bayesian posteriors and evidences},
  author={Speagle, Joshua S},
  journal={Monthly Notices of the Royal Astronomical Society},
  volume={493},
  number={3},
  pages={3132--3158},
  year={2020},
  publisher={Oxford University Press}
}

@article{farr2019accuracy,
year = {2019},
month = {may},
publisher = {The American Astronomical Society},
volume = {3},
number = {5},
pages = {66},
author = {Farr, Will M.},
title = {Accuracy Requirements for Empirically Measured Selection Functions},
journal = {Research Notes of the AAS}
}

@article{mandel2019extracting,
  title={Extracting distribution parameters from multiple uncertain observations with selection biases},
  author={Mandel, Ilya and Farr, Will M and Gair, Jonathan R},
  journal={Monthly Notices of the Royal Astronomical Society},
  volume={486},
  number={1},
  pages={1086--1093},
  year={2019},
  publisher={Oxford University Press}
}

@article{chen2022confronting,
  title={Confronting the primordial black hole scenario with the gravitational-wave events detected by LIGO-Virgo},
  author={Chen, Zu-Cheng and Yuan, Chen and Huang, Qing-Guo},
  journal={Physics Letters B},
  volume={829},
  pages={137040},
  year={2022},
  publisher={Elsevier}
}

@article{liu2023constraining,
  title={Constraining the merger history of primordial-black-hole binaries from GWTC-3},
  author={Liu, Lang and You, Zhi-Qiang and Wu, You and Chen, Zu-Cheng},
  journal={Physical Review D},
  volume={107},
  number={6},
  pages={063035},
  year={2023},
  publisher={APS}
}

@article{bagui2025primordial,
  title={Primordial black holes and their gravitational-wave signatures},
  author={Bagui, Eleni and Clesse, Sebastien and De Luca, Valerio and Ezquiaga, Jose Mar{\'\i}a and Franciolini, Gabriele and Garc{\'\i}a-Bellido, Juan and Joana, Cristian and Kumar Jain, Rajeev and Kuroyanagi, Sachiko and Musco, Ilia and others},
  journal={Living Reviews in Relativity},
  volume={28},
  number={1},
  pages={1},
  year={2025},
  publisher={Springer}
}

@book{schneider2006gravitational,
  title={Gravitational lensing: strong, weak and micro: Saas-Fee advanced course 33},
  author={Schneider, Peter and Kochanek, Christopher and Wambsganss, Joachim},
  volume={33},
  year={2006},
  publisher={Springer Science \& Business Media}
}

@article{shan2023microlensing,
  title={Microlensing sheds light on the detection of strong lensing gravitational waves},
  author={Shan, Xikai and Chen, Xuechun and Hu, Bin and Cai, Rong-Gen},
  journal={arXiv preprint arXiv:2301.06117},
  year={2023}
}

@article{chan2024detectability,
  title={Detectability of Lensed Gravitational Waves in Matched-Filtering Searches},
  author={Chan, Juno CL and Seo, Eungwang and Li, Alvin KY and Fong, Heather and Ezquiaga, Jose M},
  journal={arXiv preprint arXiv:2411.13058},
  year={2024}
}

@article{abbott2024search,
  title={Search for gravitational-lensing signatures in the full third observing run of the LIGO--Virgo network},
  author={Abbott, R and Abe, H and Acernese, F and Ackley, Kendall and Adhicary, S and Adhikari, Naresh and Adhikari, Rana X and Adkins, VK and Adya, VB and Affeldt, C and others},
  journal={The Astrophysical Journal},
  volume={970},
  number={2},
  pages={191},
  year={2024},
  publisher={The American Astronomical Society}
}

@article{hannuksela2019search,
  title={Search for gravitational lensing signatures in LIGO-Virgo binary black hole events},
  author={Hannuksela, OA and Haris, K and Ng, KKY and Kumar, S and Mehta, AK and Keitel, D and Li, TGF and Ajith, P},
  journal={The Astrophysical Journal Letters},
  volume={874},
  number={1},
  pages={L2},
  year={2019},
  publisher={IOP Publishing}
}

@article{cheung2021stellar,
  title={Stellar-mass microlensing of gravitational waves},
  author={Cheung, Mark HY and Gais, Joseph and Hannuksela, Otto A and Li, Tjonnie GF},
  journal={Monthly Notices of the Royal Astronomical Society},
  volume={503},
  number={3},
  pages={3326--3336},
  year={2021},
  publisher={Oxford University Press}
}

@article{PhysRevD.98.083005,
  title = {Discovering intermediate-mass black hole lenses through gravitational wave lensing},
  author = {Lai, Kwun-Hang and Hannuksela, Otto A. and Herrera-Mart\'{\i}n, Antonio and Diego, Jose M. and Broadhurst, Tom and Li, Tjonnie G. F.},
  journal = {Phys. Rev. D},
  volume = {98},
  issue = {8},
  pages = {083005},
  numpages = {7},
  year = {2018},
  month = {Oct},
  publisher = {American Physical Society},
  doi = {10.1103/PhysRevD.98.083005},
  url = {https://link.aps.org/doi/10.1103/PhysRevD.98.083005}
}

@article{PhysRevD.98.103022,
  title = {Detecting stellar lensing of gravitational waves with ground-based observatories},
  author = {Christian, Pierre and Vitale, Salvatore and Loeb, Abraham},
  journal = {Phys. Rev. D},
  volume = {98},
  issue = {10},
  pages = {103022},
  numpages = {7},
  year = {2018},
  month = {Nov},
  publisher = {American Physical Society},
  doi = {10.1103/PhysRevD.98.103022},
  url = {https://link.aps.org/doi/10.1103/PhysRevD.98.103022}
}

@article{diego2019observational,
  title={Observational signatures of microlensing in gravitational waves at LIGO/Virgo frequencies},
  author={Diego, Jos{\'e} Mar{\'\i}a and Hannuksela, Otto A and Kelly, Patrick L and Pagano, G and Broadhurst, Tom and Kim, K and Li, Tjonnie Guang Feng and Smoot, George Fitzgerald},
  journal={Astronomy \& Astrophysics},
  volume={627},
  pages={A130},
  year={2019},
  publisher={EDP Sciences}
}

@article{madau2014cosmic,
  title={Cosmic star-formation history},
  author={Madau, Piero and Dickinson, Mark},
  journal={Annual Review of Astronomy and Astrophysics},
  volume={52},
  number={1},
  pages={415--486},
  year={2014},
  publisher={Annual Reviews}
}

@article{PhysRevD.101.123512,
  title = {Constraining the abundance of primordial black holes with gravitational lensing of gravitational waves at LIGO frequencies},
  author = {Diego, Jose M.},
  journal = {Phys. Rev. D},
  volume = {101},
  issue = {12},
  pages = {123512},
  numpages = {19},
  year = {2020},
  month = {Jun},
  publisher = {American Physical Society},
  doi = {10.1103/PhysRevD.101.123512},
  url = {https://link.aps.org/doi/10.1103/PhysRevD.101.123512}
}

@article{mishra2021gravitational,
  title={Gravitational lensing of gravitational waves: effect of microlens population in lensing galaxies},
  author={Mishra, Anuj and Meena, Ashish Kumar and More, Anupreeta and Bose, Sukanta and Bagla, Jasjeet Singh},
  journal={Monthly Notices of the Royal Astronomical Society},
  volume={508},
  number={4},
  pages={4869--4886},
  year={2021},
  publisher={Oxford University Press}
}

@article{seo2022improving,
  title={Improving Detection of Gravitational-wave Microlensing Using Repeated Signals Induced by Strong Lensing},
  author={Seo, Eungwang and Hannuksela, Otto A and Li, Tjonnie GF},
  journal={The Astrophysical Journal},
  volume={932},
  number={1},
  pages={50},
  year={2022},
  publisher={IOP Publishing}
}

@article{yeung2023detectability,
  title={Detectability of microlensed gravitational waves},
  author={Yeung, Simon MC and Cheung, Mark HY and Seo, Eungwang and Gais, Joseph AJ and Hannuksela, Otto A and Li, Tjonnie GF},
  journal={Monthly Notices of the Royal Astronomical Society},
  volume={526},
  number={2},
  pages={2230--2240},
  year={2023},
  publisher={Oxford University Press}
}

@article{meena2022gravitational,
  title={Gravitational lensing of gravitational waves: Probability of microlensing in galaxy-scale lens population},
  author={Meena, Ashish Kumar and Mishra, Anuj and More, Anupreeta and Bose, Sukanta and Bagla, Jasjeet Singh},
  journal={Monthly Notices of the Royal Astronomical Society},
  volume={517},
  number={1},
  pages={872--884},
  year={2022},
  publisher={Oxford University Press}
}

@article{Shan:2023qvd,
    author = "Shan, Xikai and Chen, Xuechun and Hu, Bin and Li, Guoliang",
    title = "{Microlensing bias on the detection of strong lensing gravitational wave}",
    eprint = "2306.14796",
    archivePrefix = "arXiv",
    primaryClass = "astro-ph.CO",
    doi = "10.1007/s11433-023-2334-9",
    journal = "Sci. China Phys. Mech. Astron.",
    volume = "67",
    number = "6",
    pages = "269511",
    year = "2024"
}

@article{seo2025residual,
  title={Residual test to search for microlensing signatures in strongly lensed gravitational wave signals},
  author={Seo, Eungwang and Shan, Xikai and Janquart, Justin and Hannuksela, Otto and Hendry, Martin and Hu, Bin},
  journal={The Astrophysical Journal},
  volume={988},
  number={2},
  pages={159},
  year={2025},
  publisher={The American Astronomical Society}
}

@article{cardoso2019testing,
  title={Testing the nature of dark compact objects: a status report: V. Cardoso, P. Pani},
  author={Cardoso, Vitor and Pani, Paolo},
  journal={Living Reviews in Relativity},
  volume={22},
  number={1},
  pages={4},
  year={2019},
  publisher={Springer}
}

@article{zumalacarregui2018limits,
  title={Limits on stellar-mass compact objects as dark matter from gravitational lensing of type Ia supernovae},
  author={Zumalacarregui, Miguel and Seljak, Uro{\v{s}}},
  journal={Physical review letters},
  volume={121},
  number={14},
  pages={141101},
  year={2018},
  publisher={APS}
}

@article{green2026stellar,
  title={Stellar microlensing surveys as a probe of Primordial Black Holes: status and prospects},
  author={Green, Anne M},
  journal={arXiv preprint arXiv:2602.15974},
  year={2026}
}

@article{alcock2001macho,
  title={MACHO Project Limits on Black Hole Dark Matter in the 1-30 M☉ Range},
  author={Alcock, Charles and Allsman, RA and Alves, David R and Axelrod, TS and Becker, Andrew C and Bennett, DP and Cook, Kem H and Dalal, N and Drake, Andrew John and Freeman, KC and others},
  journal={The Astrophysical Journal Letters},
  volume={550},
  number={2},
  pages={L169--L172},
  year={2001}
}

@article{winch2022using,
  title={Using LSST microlensing to constrain dark compact objects in spherical and disk configurations},
  author={Winch, Harrison and Setford, Jack and Bovy, Jo and Curtin, David},
  journal={The Astrophysical Journal},
  volume={933},
  number={2},
  pages={177},
  year={2022},
  publisher={The American Astronomical Society}
}

@article{fardeen2024astrometric,
  title={Astrometric microlensing by primordial black holes with the roman space telescope},
  author={Fardeen, James and McGill, Peter and Perkins, Scott E and Dawson, William A and Abrams, Natasha S and Lu, Jessica R and Ho, Ming-Feng and Bird, Simeon},
  journal={The Astrophysical Journal},
  volume={965},
  number={2},
  pages={138},
  year={2024},
  publisher={The American Astronomical Society}
}

@article{romao2025dark,
  title={Dark classification matters: searching for primordial black holes with LSST},
  author={Romao, Miguel Crispim and Croon, Djuna and Crossey, Benedict and Godines, Daniel},
  journal={Journal of Cosmology and Astroparticle Physics},
  volume={2025},
  number={10},
  pages={066},
  year={2025},
  publisher={IOP Publishing}
}

@article{niikura2019microlensing,
  title={Microlensing constraints on primordial black holes with Subaru/HSC Andromeda observations},
  author={Niikura, Hiroko and Takada, Masahiro and Yasuda, Naoki and Lupton, Robert H and Sumi, Takahiro and More, Surhud and Kurita, Toshiki and Sugiyama, Sunao and More, Anupreeta and Oguri, Masamune and others},
  journal={Nature Astronomy},
  volume={3},
  number={6},
  pages={524--534},
  year={2019},
  publisher={Nature Publishing Group UK London}
}

@article{mroz2019microlensing,
  title={Microlensing optical depth and event rate toward the galactic bulge from 8 yr of OGLE-IV observations},
  author={Mr{\'o}z, Przemek and Udalski, Andrzej and Skowron, Jan and Szyma{\'n}ski, Micha{\l} K and Soszy{\'n}ski, Igor and Wyrzykowski, {\L}ukasz and Pietrukowicz, Pawe{\l} and Koz{\l}owski, Szymon and Poleski, Rados{\l}aw and Ulaczyk, Krzysztof and others},
  journal={The Astrophysical Journal Supplement Series},
  volume={244},
  number={2},
  pages={29},
  year={2019},
  publisher={The American Astronomical Society}
}

@article{abrams2020hunting,
  title={Hunting gravitational wave black holes with microlensing},
  author={Abrams, Natasha S and Takada, Masahiro},
  journal={The Astrophysical Journal},
  volume={905},
  number={2},
  pages={121},
  year={2020},
  publisher={The American Astronomical Society}
}

@article{oguri2018understanding,
  title={Understanding caustic crossings in giant arcs: characteristic scales, event rates, and constraints on compact dark matter},
  author={Oguri, Masamune and Diego, Jose M and Kaiser, Nick and Kelly, Patrick L and Broadhurst, Tom},
  journal={Physical Review D},
  volume={97},
  number={2},
  pages={023518},
  year={2018},
  publisher={APS}
}

@article{blaineau2022new,
  title={New limits from microlensing on Galactic black holes in the mass range 10 M⊙< M< 1000 M⊙},
  author={Blaineau, Tristan and Moniez, M and Afonso, C and Albert, J-N and Ansari, R and Aubourg, E and Coutures, C and Glicenstein, J-F and Goldman, B and Hamadache, C and others},
  journal={Astronomy \& Astrophysics},
  volume={664},
  pages={A106},
  year={2022},
  publisher={EDP Sciences}
}

@article{perkins2024disentangling,
  title={Disentangling the black hole mass spectrum with photometric microlensing surveys},
  author={Perkins, Scott E and McGill, Peter and Dawson, William and Abrams, Natasha S and Lam, Casey Y and Ho, Ming-Feng and Lu, Jessica R and Bird, Simeon and Pruett, Kerianne and Golovich, Nathan and others},
  journal={The Astrophysical Journal},
  volume={961},
  number={2},
  pages={179},
  year={2024},
  publisher={The American Astronomical Society}
}

@article{zhang2023dark,
  title={Dark stars and gravitational waves: Topical review},
  author={Zhang, Kilar and Luo, Ling-Wei and Tsao, Jie-Shiun and Chen, Chian-Shu and Lin, Feng-Li},
  journal={Results in Physics},
  volume={53},
  pages={106967},
  year={2023},
  publisher={Elsevier}
}

@article{green2021primordial,
  title={Primordial Black Holes as a dark matter candidate},
  author={Green, Anne M and Kavanagh, Bradley J},
  journal={Journal of Physics G: Nuclear and Particle Physics},
  volume={48},
  number={4},
  pages={043001},
  year={2021},
  publisher={IOP Publishing}
}

@incollection{carr2025history,
  title={The history of primordial black holes},
  author={Carr, Bernard J and Green, Anne M},
  booktitle={Primordial black holes},
  pages={3--33},
  year={2025},
  publisher={Springer}
}

@ARTICLE{2026arXiv260527226T,
       author = {{The LIGO Scientific Collaboration} and {the Virgo Collaboration} and {the KAGRA Collaboration} and {Abac}, A.~G. and {Abe}, A. and {Abouelfettouh}, I. and {Acernese}, F. and {Ackley}},
        title = "{GWTC-5.0: Population Properties of Merging Compact Binaries}",
      journal = {arXiv e-prints},
         year = {2026},
         
archivePrefix = {arXiv},
       eprint = {2605.27226},
 primaryClass = {astro-ph.HE}
}

@article{diemand2008clumps,
  title={Clumps and streams in the local dark matter distribution},
  author={Diemand, J{\"u}rg and Kuhlen, Michael and Madau, Piero and Zemp, M and Moore, Ben and Potter, D and Stadel, J},
  journal={Nature},
  volume={454},
  number={7205},
  pages={735--738},
  year={2008},
  publisher={Nature Publishing Group UK London}
}

@article{fedorova2016gravitational,
  title={Gravitational microlensing as a probe for dark matter clumps},
  author={Fedorova, E and Sliusar, VM and Zhdanov, VI and Alexandrov, AN and Popolo, A Del and Surdej, Jean},
  journal={Monthly Notices of the Royal Astronomical Society},
  volume={457},
  number={4},
  pages={4147--4159},
  year={2016},
  publisher={The Royal Astronomical Society}
}

@article{silk1993clumpy,
  title={Clumpy cold dark matter},
  author={Silk, Joseph and Stebbins, Albert},
  journal={Astrophysical Journal-Part 1},
  volume={411},
  pages={439--449},
  year={1993}
}

@article{buckley2018collapsed,
  title={Collapsed dark matter structures},
  author={Buckley, Matthew R and DiFranzo, Anthony},
  journal={Physical review letters},
  volume={120},
  number={5},
  pages={051102},
  year={2018},
  publisher={APS}
}

@article{ryan2022exotic,
  title={Exotic compact objects: The dark white dwarf},
  author={Ryan, Michael and Radice, David},
  journal={Physical Review D},
  volume={105},
  number={11},
  pages={115034},
  year={2022},
  publisher={APS}
}

@article{d2018massive,
  title={Massive black holes from dissipative dark matter},
  author={d'Amico, Guido and Panci, Paolo and Lupi, Alessandro and Bovino, Stefano and Silk, Joe},
  journal={Monthly Notices of the Royal Astronomical Society},
  volume={473},
  number={1},
  pages={328--335},
  year={2018},
  publisher={Oxford University Press}
}

@article{chang2019structure,
  title={Structure formation and exotic compact objects in a dissipative dark sector},
  author={Chang, Jae Hyeok and Egana-Ugrinovic, Daniel and Essig, Rouven and Kouvaris, Chris},
  journal={Journal of Cosmology and Astroparticle Physics},
  volume={2019},
  number={03},
  pages={036--036},
  year={2019}
}

@article{bramante2024effect,
  title={The effect of multiple cooling channels on the formation of dark compact objects},
  author={Bramante, Joseph and Diamond, Melissa and Kim, J Leo},
  journal={Journal of Cosmology and Astroparticle Physics},
  volume={2024},
  number={02},
  pages={002},
  year={2024},
  publisher={IOP Publishing}
}

@article{mao2012astrophysical,
  title={Astrophysical applications of gravitational microlensing},
  author={Mao, Shude},
  journal={Research in Astronomy and Astrophysics},
  volume={12},
  number={8},
  pages={947--972},
  year={2012}
}

@article{gould2000natural,
  title={A natural formalism for microlensing},
  author={Gould, Andrew},
  journal={The Astrophysical Journal},
  volume={542},
  number={2},
  pages={785--788},
  year={2000}
}

@article{gould2004resolution,
  title={Resolution of the MACHO-LMC-5 Puzzle: the jerk-parallax microlens degeneracy},
  author={Gould, Andrew},
  journal={The Astrophysical Journal},
  volume={606},
  number={1},
  pages={319--325},
  year={2004}
}

@article{an2002first,
  title={First microlens mass measurement: PLANET photometry of EROS BLG-2000-5},
  author={An, Jin H and Albrow, Michael D and Beaulieu, J-P and Caldwell, JAR and DePoy, DL and Dominik, Martin and Gaudi, BS and Gould, A and Greenhill, J and Hill, K and others},
  journal={The Astrophysical Journal},
  volume={572},
  number={1},
  pages={521--539},
  year={2002}
}

@article{gould1992extending,
  title={Extending the MACHO search to about 10 exp 6 solar masses},
  author={Gould, Andrew},
  journal={Astrophysical Journal, Part 1},
  volume={392},
  pages={442--451},
  year={1992}
}

@article{dominik2000astrometric,
  title={Astrometric microlensing of stars},
  author={Dominik, Martin and Sahu, Kailash C},
  journal={The Astrophysical Journal},
  volume={534},
  number={1},
  pages={213--226},
  year={2000}
}

@article{coleman2004intermediate,
  title={Intermediate-mass black holes},
  author={Coleman Miller, M and Colbert, Edward JM},
  journal={International Journal of Modern Physics D},
  volume={13},
  number={01},
  pages={1--64},
  year={2004},
  publisher={World Scientific}
}

@article{carr2016primordial,
  title={Primordial black holes as dark matter},
  author={Carr, Bernard and K{\"u}hnel, Florian and Sandstad, Marit},
  journal={Physical Review D},
  volume={94},
  number={8},
  pages={083504},
  year={2016},
  publisher={APS}
}

@article{belczynski2004comprehensive,
  title={A comprehensive study of young black hole populations},
  author={Belczynski, Krzysztof and Sadowski, Aleksander and Rasio, Frederic A},
  journal={The Astrophysical Journal},
  volume={611},
  number={2},
  pages={1068--1079},
  year={2004}
}

@article{narayan2005black,
  title={Black holes in astrophysics},
  author={Narayan, Ramesh},
  journal={New Journal of Physics},
  volume={7},
  number={1},
  pages={199--199},
  year={2005}
}

@article{gaudi2012microlensing,
  title={Microlensing surveys for exoplanets},
  author={Gaudi, B Scott},
  journal={Annual Review of Astronomy and Astrophysics},
  volume={50},
  number={1},
  pages={411--453},
  year={2012},
  publisher={Annual Reviews}
}

@article{abac2026gwtc_catalog,
  title={GWTC-5.0: Observations from the Second Part of the Fourth LIGO-Virgo-KAGRA Observing Run and Updates to the Gravitational-Wave Transient Catalog},
  author={Abac, AG and Abe, A and Abouelfettouh, I and Acernese, F and Ackley, K and Adam, A and Adhicary, S and Adhikari, D and Adhikari, RX and Adkins, VK and others},
  journal={arXiv preprint arXiv:2605.27225},
  year={2026}
}

@article{callister2020shouts,
  title={Shouts and murmurs: Combining individual gravitational-wave sources with the stochastic background to measure the history of binary black hole mergers},
  author={Callister, Tom and Fishbach, Maya and Holz, Daniel E and Farr, Will M},
  journal={The Astrophysical Journal Letters},
  volume={896},
  number={2},
  pages={L32},
  year={2020},
  publisher={The American Astronomical Society}
}

@article{babak2013searching,
  title={Searching for gravitational waves from binary coalescence},
  author={Babak, S and Biswas, R and Brady, PR and Brown, Duncan A and Cannon, K and Capano, Collin D and Clayton, Jessica H and Cokelaer, T and Creighton, Jolien DE and Dent, T and others},
  journal={Physical Review D},
  volume={87},
  number={2},
  pages={024033},
  year={2013},
  publisher={APS}
}

@article{PhysRevD.94.063530,
  title = {Microlensing and dynamical constraints on primordial black hole dark matter with an extended mass function},
  author = {Green, Anne M.},
  journal = {Physical Review D},
  volume = {94},
  issue = {6},
  pages = {063530},
  numpages = {6},
  year = {2016},
  month = {Sep},
  publisher = {American Physical Society}
}

@article{goyal2026across,
  title={Across the Universe: GW231123 as a magnified and diffracted black hole merger},
  author={Goyal, Srashti and Villarrubia-Rojo, Hector and Zumalac{\'a}rregui, Miguel},
  journal={The Astrophysical Journal Letters},
  volume={1008},
  number={1},
  pages={L12},
  year={2026},
  publisher={The American Astronomical Society}
}

@article{ho2026diffraction,
  title={The diffraction-lensing interpretation of GW231123 with astrophysical priors},
  author={Ho-Yeuk Cheung, Mark and Wadekar, Digvijay and Zaldarriaga, Matias and Venumadhav, Tejaswi and Roulet, Javier and Mehta, Ajit Kumar},
  journal={arXiv e-prints},
  pages={arXiv--2607},
  year={2026}
}

@article{carr2017primordial,
  title={Primordial black hole constraints for extended mass functions},
  author={Carr, Bernard and Raidal, Martti and Tenkanen, Tommi and Vaskonen, Ville and Veerm{\"a}e, Hardi},
  journal={Physical Review D},
  volume={96},
  number={2},
  pages={023514},
  year={2017},
  publisher={APS}
}

@article{carr2017primomatterdom,
  title={Primordial black holes from inflaton and spectator field perturbations in a matter-dominated era},
  author={Carr, Bernard and Tenkanen, Tommi and Vaskonen, Ville},
  journal={Physical Review D},
  volume={96},
  number={6},
  pages={063507},
  year={2017},
  publisher={APS}
}

@article{james2020constraints,
  title={Constraints on the cosmic string loop collapse fraction from primordial black holes},
  author={James-Turner, Chloe and Weil, Danton PB and Green, Anne M and Copeland, Edmund J},
  journal={Physical Review D},
  volume={101},
  number={12},
  pages={123526},
  year={2020},
  publisher={APS}
}

@article{talbot2018measuring,
  title={Measuring the binary black hole mass spectrum with an astrophysically motivated parameterization},
  author={Talbot, Colm and Thrane, Eric},
  journal={The Astrophysical Journal},
  volume={856},
  number={2},
  pages={173},
  year={2018},
  publisher={The American Astronomical Society}
}

@article{chan2026discovering,
  title={Discovering gravitational waveform distortions from lensing: a deep dive into GW231123},
  author={Chan, Juno CL and Ezquiaga, Jose Mar{\'\i}a and Lo, Rico KL and Bowman, Joey and Zertuche, Lorena Maga{\~n}a and Vujeva, Luka},
  journal={Physical Review D},
  volume={114},
  number={2},
  pages={L021502},
  year={2026},
  publisher={APS}
}

@article{chakraborty2026first,
  title={The First Model-independent Upper Bound on Microlensing Signature of the Highest-mass Binary Black Hole Event GW231123},
  author={Chakraborty, Aniruddha and Mukherjee, Suvodip},
  journal={The Astrophysical Journal},
  volume={1003},
  number={1},
  pages={20},
  year={2026},
  publisher={The American Astronomical Society}
}

@article{cheung2026diffraction,
  title={The diffraction-lensing interpretation of GW231123 with astrophysical priors},
  author={Cheung, Mark Ho-Yeuk and Wadekar, Digvijay and Zaldarriaga, Matias and Venumadhav, Tejaswi and Roulet, Javier and Mehta, Ajit Kumar},
  journal={arXiv preprint arXiv:2607.21834},
  year={2026}
}

@article{hu2026gw231123,
  title={GW231123: Overlapping Gravitational Wave Signals?},
  author={Hu, Qian and Narola, Harsh and Heynen, Jef and Wright, Mick and Veitch, John and Janquart, Justin and Van Den Broeck, Chris},
  journal={The Astrophysical Journal},
  volume={1003},
  number={2},
  pages={187},
  year={2026},
  publisher={The American Astronomical Society}
}

@article{shan2025gw231123,
  title={GW231123: A case for binary microlensing in a strong lensing field},
  author={Shan, Xikai and Yang, Huan and Mao, Shude},
  journal={arXiv preprint arXiv:2512.19118},
  year={2025}
}

@article{wang2026gw231123,
  title={GW231123: False Massive Graviton Signatures from Unmodeled Point-Mass Lensing},
  author={Wang, Baoxiang and Yang, Tao},
  journal={arXiv preprint arXiv:2604.08179},
  year={2026}
}

@article{abbott2021search,
  title={Search for lensing signatures in the gravitational-wave observations from the first half of LIGO--Virgo’s third observing run},
  author={Abbott, Richard and Abbott, Thomas D and Abraham, S and Acernese, Fausto and Ackley, K and Adams, A and Adams, C and Adhikari, Rana X and Adya, VB and Affeldt, C and others},
  journal={The Astrophysical Journal},
  volume={923},
  number={1},
  pages={14},
  year={2021},
  publisher={The American Astronomical Society}
}

@article{capote2025advanced,
  title={Advanced LIGO detector performance in the fourth observing run},
  author={Capote, E and Jia, W and Aritomi, N and Nakano, M and Xu, V and Abbott, R and Abouelfettouh, I and Adhikari, RX and Ananyeva, A and Appert, S and others},
  journal={Physical Review D},
  volume={111},
  number={6},
  pages={062002},
  year={2025},
  publisher={APS}
}

@article{acernese2026advanced,
  title={Advanced Virgo during the LIGO-Virgo-KAGRA fourth observing run},
  author={Acernese, F and Agapito, A and Agarwal, D and Ahrend, IL and Aiello, L and Ain, A and Ali, W and Allocca, A and Amar, W and Amato, A and others},
  journal={arXiv preprint arXiv:2607.26872},
  year={2026}
}
\bibliographystyle{aasjournalv7}

\appendix

\section{Mock BBH Population}\label{app:source_pop}
We generate the population of simulated unlensed source BBHs from the \textsc{PowerLaw+Peak} primary mass model and a Madau--Dickinson-like \citep{madau2014cosmic} redshift distribution, adopting the posterior median values of the population hyperparameters inferred from GWTC-3~\citep{abbott2023population}, as summarized in Table~\ref{tab:hyperparameters}.

\begin{table}[h]
\centering
\begin{tabular}{l l l}
\hline\hline
Parameter & Description & Injected value \\
\hline
$\alpha$      & Spectral index for the power law of the primary mass distribution & 3.78 \\
$\beta$     & Spectral index for the power law of the mass ratio distribution & 0.81 \\
$m_{\min}$    & Minimum mass of the power-law component of the black hole mass distribution & 4.98$M_{\odot}$ \\
$m_{\max}$    & Maximum mass of the power-law component of the black hole mass distribution & 112.5$M_{\odot}$ \\
$\lambda_{\rm g}$ & Fraction of BBH systems in the Gaussian component & 0.03 \\
$\mu_{g}$  & Mean of the Gaussian component in the primary mass distribution & 32.27$M_\odot$ \\
$\sigma_{g}$    & Standard deviation of the Gaussian component in the primary mass distribution & 3.88$M_\odot$ \\
$\delta_m$    & Range of mass tapering at the lower end of the mass distribution & 4.8$M_{\odot}$ \\
$z_{p}$    & Redshift at which the distribution peaks & 2.7 \\
$\gamma$   & Power-law index of the redshift distribution for $z < z_p$ & 4.59 \\
$\kappa$    &  Power-law index of the redshift distribution for $z > z_p$& 2.86 \\
\hline
\end{tabular}
\caption{List of hyperparameters of the BBH population model and their injected values used in our GWTC-3-based simulations.}
\label{tab:hyperparameters}
\end{table}
\subsection{Mass distribution}\label{app:component_mass}

The primary source-frame mass $m_1$ is drawn from a truncated power law  defined over $m_{\rm min}= 5M_{\odot}$ and $m_{\rm max}=112M_{\odot}$, with a Gaussian peak and a smooth low-mass taper:
\begin{equation}
\label{eq:m1_pop}
p_{\rm pop}\left(m_1 \mid  \alpha, \lambda_{\mathrm{g}}, \mu_{\mathrm{g}}, \sigma_{\mathrm{g}}, m_{\rm min}, \delta_{m}\right)=\left[\left(1-\lambda_{\mathrm{g}}\right) m^{-\alpha}_{1}+ 
\frac{\lambda_{\mathrm{g}}}{\sqrt{2\pi}\,\sigma_g}
\exp\left(
-\frac{(m_1 - \mu_g)^2}{2\sigma_g^2}
\right)\right]S(m_{1}|m_{\rm min}, \delta_{m}),
\end{equation}
where $\alpha$ denotes the power-law index, and $\lambda_{g}$, $\mu_{g}$, and $\sigma_{g}$ correspond to the mixture fraction, mean, and standard deviation of the Gaussian component, respectively.
$S(m_{1}|m_{\rm min},\delta_{m})$ represents a smoothing function, which is given by
\begin{equation}
    \label{eq:smoothing}
    \begin{aligned}
& S\left(m \mid m_{\min }, \delta_m\right) \\
& \quad= \begin{cases}0 & \left(m<m_{\min }\right), \\
{\left[\mathrm{exp}\left(\frac{\delta_{m}}{m-m_{\rm min}}+\frac{\delta_{m}}{m-m_{\rm{min}}-\delta_{m}}\right)+1\right]^{-1}} & \left(m_{\min } \leq m<m_{\min }+\delta_m\right), \\
1 & \left(m \geq m_{\min }+\delta_m\right),\end{cases}
\end{aligned}
\end{equation}
where $\delta_{m}$ sets the width of the low-mass taper applied near $m_{\rm min}$.

The secondary mass $m_2$ is determined by the mass ratio $q\equiv m_2/m_1$, which is drawn from a truncated power-law distribution with the low-mass smoothing function defined in Eq.~\eqref{eq:smoothing}:
\begin{equation}\label{eq:qm_pop}
p_{\rm pop}(q\mid \beta, m_1,m_{\min},\delta_m)\propto q^{\beta}\,S(q m_1\mid m_{\min},\delta_m),\qquad m_{\min}<m_2\le m_1,
\end{equation}
where $\beta$ denotes the power-law index of the $q$ distribution.
\subsection{Redshift distribution}\label{app:redshift}

For a source-frame merger-rate density $\mathcal{R}(z)$ at redshift $z$, the number of mergers per unit detector-frame time and redshift is given by:
\begin{equation}\label{eq:diff_merger_rate}
\frac{d^{2}N}{dt\,dz}=\frac{\mathcal{R}(z)}{1+z}\frac{dV_c}{dz},
\end{equation}
where $V_c$ is the comoving volume at $z$.
Following the parametrization of \citet{callister2020shouts}, which builds on the star-formation-rate form of \citet{madau2014cosmic}, the merger rate density can be written as:
\begin{equation}
\label{eq:z_pop}
\mathcal{R}(z) = \mathcal{R}_0(1+z)^\gamma \frac{1+\left(1+z_p\right)^{-(\gamma+\kappa)}}{1+\left(\frac{1+z}{1+z_p}\right)^{\gamma+\kappa}},
\end{equation}
where $\mathcal{R}_0=\mathcal{R}(0)$, $z_{p}$ is the redshift at which $\mathcal{R}(z)$ peaks, and $\gamma$ and $\kappa$ denote the asymptotic power-law indices $z\ll z_p$ and $z\gg z_p$, respectively.
In our simulations, we adopt the local merger rate density $R_{0}=20~\mathrm{Gpc^{-3}yr^{-1}}$.
\subsection{Consistency with GWTC-5}\label{app:gwtc3vs5}
\begin{figure*}[t]
    \centering
\includegraphics[width=1.0\linewidth]{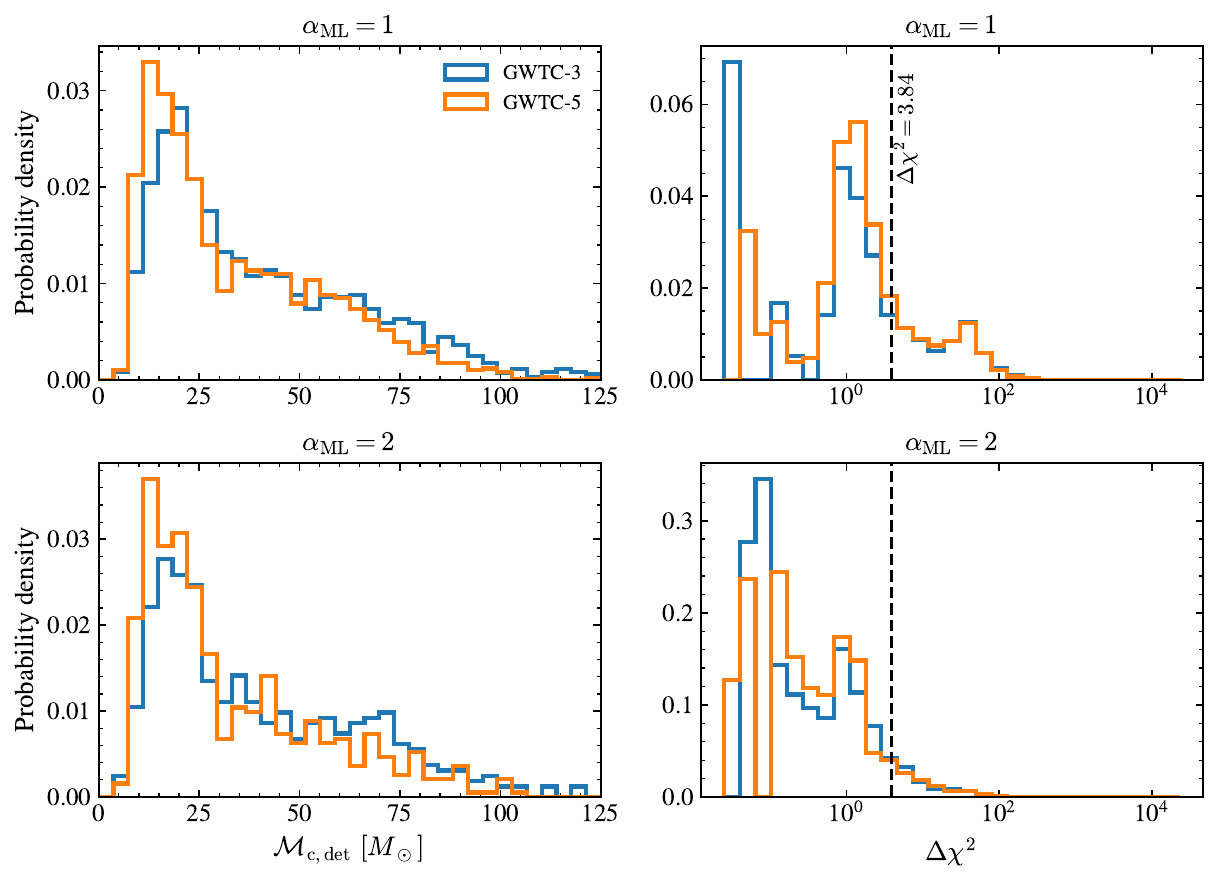}
    \caption{Comparison of the detectable microlensed BBH populations for $\alpha_{\rm ML}=1$ (top) and $\alpha_{\rm ML}=2$ (bottom) generated from the GWTC-3 (blue) and GWTC-5 (orange) population models.
    The left and right panels show the distributions of $\mathcal{M}_{\rm c,det}$ and $\Delta\chi^2$, respectively.}\label{fig:gwtc3vs5}
\end{figure*}

The most recent BBH population model is based on GWTC-5~\citep{abac2026gwtc_catalog}. 
Since repeating the full parameter-estimation analysis with the updated population is computationally expensive, we instead repeat the mock-catalog generation using the posterior-median hyperparameters from the GWTC-5 population analysis~\citep{2026arXiv260527226T}, which adopts a \textsc{Broken Power Law + 2 Peaks} model for the primary-mass distribution~\citep{talbot2018measuring,2026arXiv260527226T} and a simple power law for the redshift distribution, while retaining the same mass-ratio distribution as that given in Eq.~\eqref{eq:qm_pop}.

Relative to the GWTC-3 population, the GWTC-5 population yields a larger total number of simulated events satisfying $\rho_{\rm net}>8$ owing to its higher detector sensitivity, and therefore a larger absolute number of microlensed events. 
To compare the microlensing distinguishability on an equal footing, we instead consider the fraction of detectable microlensed events that satisfy $\Delta\chi^2>3.84$.
For $\alpha_{\rm ML}=1$ ($2$), this fraction changes from $1727/1914=90.2\%$ ($291/446=65.2\%$) for GWTC-3 to $2223/2543=87.4\%$ ($342/522=65.5\%$) for GWTC-5.
We further compare the distributions of the detector-frame chirp mass ($\mathcal{M}_{\rm c,det}$) and $\Delta\chi^2$, which are directly related to the microlensing distinguishability. 
As shown in Figure~\ref{fig:gwtc3vs5}, the GWTC-3 and GWTC-5 populations predict nearly identical distributions of these observable quantities.
The choice of BBH population model is therefore unlikely to have a significant impact on our inference of $\lambda_l$.

\begin{figure*}[t]
    \centering
\includegraphics[width=1.0\linewidth]{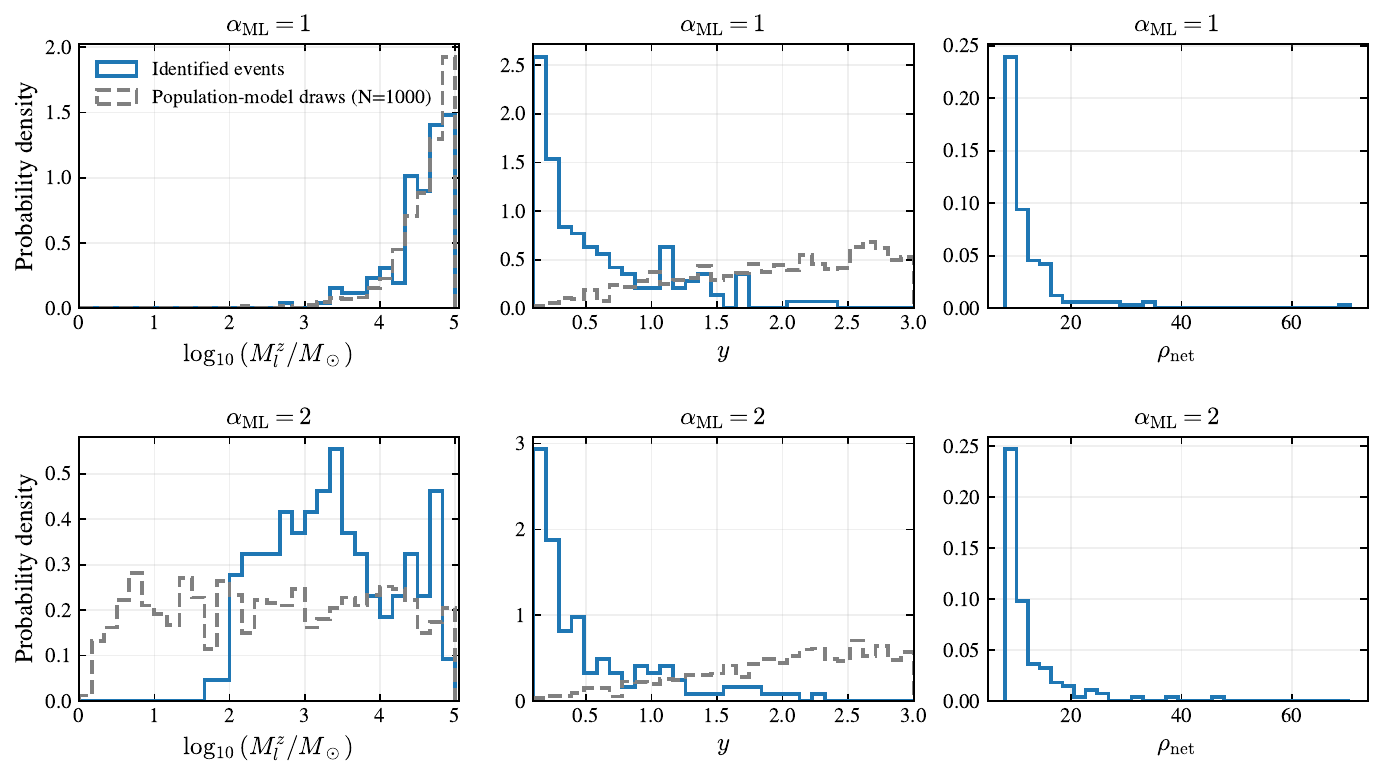}
    \caption{Histograms of the posterior medians of \(M_l^z\) and \(y\), and $\rho_{\rm net}$ of identified microlensed events with \(\mathcal{B}_{\rm ML}^{\rm U}>1\).
    The upper and lower rows correspond to \(\alpha_{\rm ML}=1\) and 2, respectively, with \(n_0=10^6\,\mathrm{Mpc}^{-3}\) fixed.
    For \(M_l^z\) and \(y\), the histograms are compared with those of 1000 samples drawn from the corresponding population models.}\label{fig:mlgw_catalog}
\end{figure*}
\section{Microlensing Optical Depths}~\label{app:tau}
Only a fraction of GW signals are expected to undergo detectable microlensing, with the microlensing probability characterized by the optical depth, $\tau(z_s)$, for a GW source at redshift $z_s$.
Assuming that the lenses are uniformly distributed, for a given $n_0$ value, the optical depth can be written as:
\begin{equation}\label{eq:lensing_optical_depth_micro}
    \tau(z_s)=
\int_0^{z_s} dz_l \int dM_l\,
n_0 p(M_l)\,
\pi y_{\max}^2 R_E^2\,
\frac{1}{4\pi D_l^2}
\frac{dV_c}{dz_l},
\end{equation}
where $p(M_{l})$ is the distribution of $M_l$ given by Eq.~\eqref{ppopml}, assuming a fixed $\alpha_{\rm ML}$ over the mass range [$M_{\rm min}$, $M_{\rm max}$].
The factor $y_{\rm max}$ is introduced because, in the wave-optics regime, diffraction effects allow detectable microlensing signatures to extend beyond $R_E$ commonly associated with EM microlensing ($y_{\rm max}=1$) (see e.g.~\cite{Caliskan:2023zqm}).
Consequently, the effective lensing cross-section expands to $\sigma = \pi y_{\rm max}^2 R_E^2$~\citep{takahashi2003wave}.
As discussed in Section~\ref{sec:2.3} (see Figure~\ref{fig:MM_dist}), we adopt $y_{\rm max}=2.75$ and $y_{\rm max}=2.07$ for the $\alpha_{\rm ML}=1$ the $\alpha_{\rm ML}=2$ populations, respectively.

The population-averaged optical depth is obtained by averaging Eq.~\eqref{eq:lensing_optical_depth_micro} over the simulated BBH redshift distribution:
\begin{equation}\label{eq:combined_optical_depth}
    \bar{\tau} = \int_{0}^{z_{\rm max}} \tau(z_{s})p_{\rm src}(z_{s})dz_{s},
\end{equation}
where $p_{\rm src}(z_s)$ is the normalized source-redshift distribution and $z_{\rm max}$ denotes the maximum source redshift considered in the simulation.
For a survey containing $N_{\rm UL}$ GW sources, the expected number of microlensed events is given by:
\begin{equation}\label{eq:total_ML}
N_{\rm ML} = N_{\rm UL} \, \bar{\tau}.
\end{equation}

\section{Lens parameters of identified microlensed events}\label{app:mlgw_catalog}
The lens properties of identified microlensed events can provide an early indication of the underlying lens-mass distribution.
Figure~\ref{fig:mlgw_catalog} compares the posterior medians of $M_l^z$ and $y$, together with the network SNRs, of events identified with $\mathcal{B}^{\rm ML}_{\rm U}>1$.
For each $\alpha_{\rm ML}$ scenario, we also show the corresponding distributions for 1000 randomly selected events from the initial $N_{\rm ML}$ simulated population. 

The two scenarios differ most notably in their $M_l^z$ distributions. 
For $\alpha_{\rm ML}=1$, events with $M_l^z<10^3~M_\odot$ are rare in the initial population and are consequently unlikely to be identified.
By contrast, the $\alpha_{\rm ML}=2$ scenario yields a broader distribution with a substantially larger fraction of low-mass lenses, making such events more likely to be identified.
Repeated detections of microlensed events with low $M_l^z$ could therefore provide an early indication in favor of $\alpha_{\rm ML}=2$.
The hierarchical Bayesian analysis quantifies this distinction by incorporating the information carried by individual events into the population-level inference.
\section{Mass-dependent dark matter fraction}\label{app:diff_fdm}
\begin{figure*}[h]
    \centering
\includegraphics[width=1.0\linewidth]{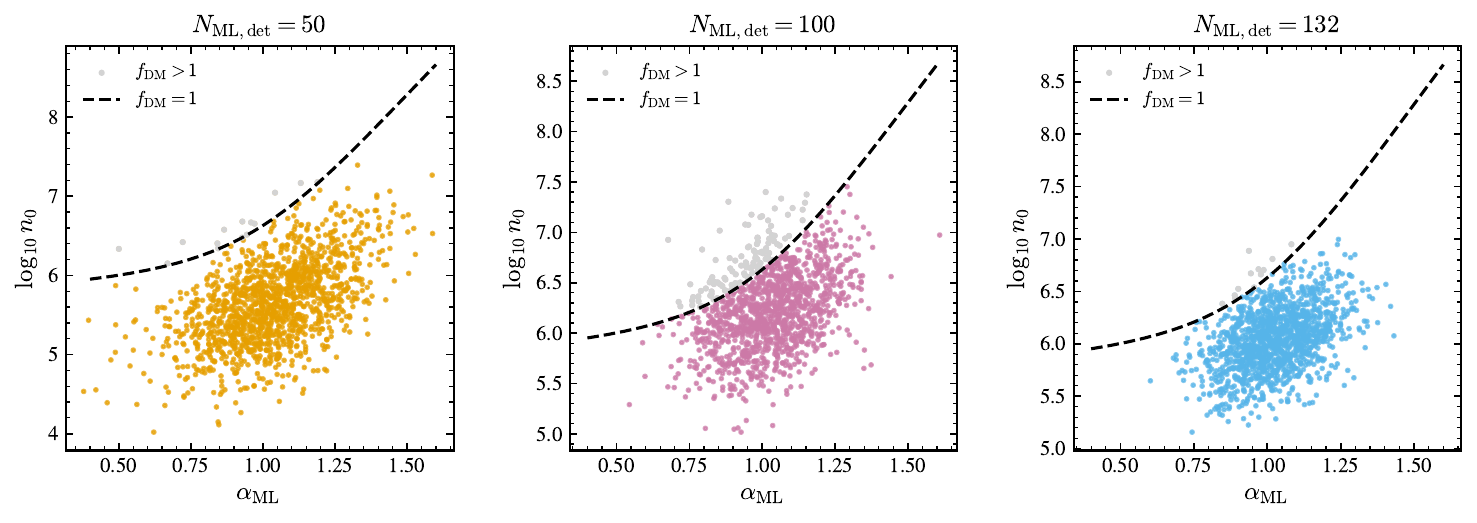}
    \caption{Joint posterior samples in the $(\alpha_{\rm ML}, n_0)$ plane for $N_{\rm ML,det}=50,100$, and $132$. The gray dots indicate the unphysical domain $f_{\rm DM}>1$ and are discarded in evaluating the differential $f_{\rm DM}$.}\label{fig:possible}
\end{figure*}

$f_{\rm DM}$ in Eq.~\eqref{eq:fdm} compresses the lens population into a single number, and in our hierarchical model it is constrained only through the correlated pair $(\alpha_{\rm ML}, n_0)$.
We therefore map the joint posterior onto the differential DM fraction that (i) provides a direct view of which lens-mass scales contribute to the inferred compact-object mass density, and (ii) allows direct comparison with mass-dependent constraints on compact-object scenarios whose abundance is expected to be concentrated within specific mass ranges, such as PBHs~\citep{carr2016primordial}.
The differential DM fraction per $\log_{10} M_l$ is given by:
\begin{equation}\label{eq:diff_fdm}
\frac{df_{\rm DM}}{d\log_{10} M_l} = \frac{8\pi G n_0}{3 H_0^2 \Omega_{\rm DM}} M_l^{2-\alpha_{\rm ML}} \ln(10).
\end{equation}

We evaluate Eq.~\eqref{eq:diff_fdm} for each joint posterior sample of $(\alpha_{\rm ML},n_0)$.
As discussed in Section~\ref{sec:4.2}, in the $\alpha_{\rm ML}=1$ scenario, a fraction of the joint posterior lies in the unphysical region $f_{\rm DM}>1$.
Figure~\ref{fig:possible} illustrates the corresponding unphysical posterior samples, which will be discarded in evaluating Eq.~\eqref{eq:diff_fdm}.

\begin{figure*}[t]
    \centering
    \includegraphics[width=1.0\linewidth]{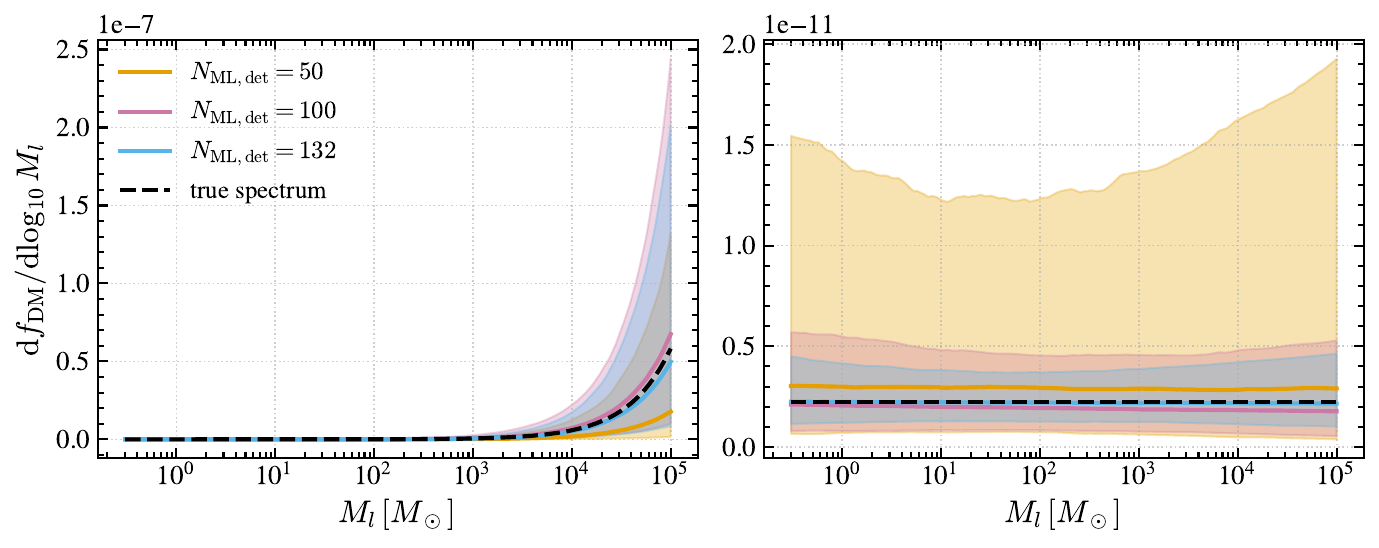}
    \caption{The differential fraction of dark matter, ${\mathrm{d}f_{\rm DM}}/{\mathrm{d}\log_{10}M_l}$, for the $\alpha_{\rm ML}=1$ (left) and $\alpha_{\rm ML}=2$ (right) scenarios, from catalogs with $N_{\rm ML,det}=50$, $100$, and $132$ detected events.
    The black dashed lines denote the corresponding true spectra, while the colored shaded regions indicate the 2-$\sigma$ credible regions after imposing $f_{\rm DM}\leq1$.}\label{fig:fdm_diff}
\end{figure*}

Figure~\ref{fig:fdm_diff} shows the differential spectra obtained from the remaining physical samples.
For $\alpha_{\rm ML}=1$, the differential DM fraction increases approximately as $M_l^{2-\alpha_{\rm ML}}\sim M_l$, so uncertainties in $\alpha_{\rm ML}$ increasingly broaden the spectrum toward higher $M_l$.
Note that the $f_{\rm DM}\leq1$ truncation preferentially removes the high-$n_0$ tail, particularly for $N_{\rm ML,det}=100$, shifting the median and narrowing the credible interval so that its constraint can appear nearly as tight as that for $N_{\rm ML,det}=132$ despite the smaller number of detected events.
This apparent tightening is caused by the physical truncation rather than by additional constraining power from the data. 

For $\alpha_{\rm ML}=2$, by contrast, the differential spectrum is substantially flatter.
In this case, ${\rm d}f_{\rm DM}/{\rm d}\log_{10}M_l$ is approximately independent of $M_l$, reducing the propagation of uncertainty in $\alpha_{\rm ML}$ toward higher $M_l$.
Reflecting the improved constraining power of the larger event sample, $N_{\rm ML,det}=132$ yields the narrowest credible interval and the closest agreement with the true spectrum among the three cases.

\end{document}